%% file: main.tex
\documentclass[11pt]{article}

\date{}

\usepackage{gensymb}
\usepackage{textcomp}
\usepackage{adjustbox}
\usepackage[utf8]{inputenc} % allow utf-8 input
\usepackage[T1]{fontenc}    % use 8-bit T1 fonts
\usepackage{url}            % simple URL typesetting
\usepackage{booktabs}       % professional-quality tables
\usepackage{amsfonts}       % blackboard math symbols
\usepackage{nicefrac}       % compact symbols for 1/2, etc.
\usepackage{microtype,nicefrac}      % microtypography
\usepackage{xspace}
\usepackage{natbib}
\usepackage{comment}
\usepackage{graphicx}
\usepackage{rotating}
\usepackage{amssymb}
\usepackage{autobreak}
\usepackage{mathtools}
\usepackage[dvipsnames]{xcolor}
\usepackage{bbm}
\usepackage[ruled,vlined, linesnumbered]{algorithm2e}
\usepackage{algorithmic}
\usepackage[parfill]{parskip}
\usepackage{color}
\usepackage{amsmath} % provides the "align" environment
\usepackage{graphicx}
\usepackage{subfig}
\usepackage{xr}
\usepackage[urlcolor=blue,citecolor=black,linkcolor=black]{hyperref}
\allowdisplaybreaks

\title{Ambient Noise Full Waveform Inversion with Neural Operators}
\author{Caifeng Zou$^1$\thanks{Corresponding author: czou@caltech.edu}, Zachary E. Ross$^1$, Robert W. Clayton$^1$,\\Fan-Chi Lin$^2$, and Kamyar Azizzadenesheli$^3$ \\\\
$^1$ \textit{Seismological Laboratory, California Institute of Technology, Pasadena, CA \emph{91125}, USA}\\
$^2$ \textit{Department of Geology and Geophysics, University of Utah, Salt Lake City, UT \emph{84112}, USA}\\
$^3$ \textit{Nvidia Corporation, Santa Clara, CA \emph{95051}, USA}
  }

\begin{document}

\label{firstpage}

\maketitle

\begin{abstract}
Numerical simulations of seismic wave propagation are crucial for investigating velocity structures and improving seismic hazard assessment. However, standard methods such as finite difference or finite element are computationally expensive. Recent studies have shown that a new class of machine learning models, called neural operators, can solve the elastodynamic wave equation orders of magnitude faster than conventional methods. Full waveform inversion is a prime beneficiary of the accelerated simulations. Neural operators, as end-to-end differentiable operators, combined with automatic differentiation, provide an alternative approach to the adjoint-state method. State-of-the-art optimization techniques built into PyTorch provide neural operators with greater flexibility to improve the optimization dynamics of full waveform inversion, thereby mitigating cycle-skipping problems. In this study, we demonstrate the first application of neural operators for full waveform inversion on a real seismic dataset, which consists of several nodal transects collected across the San Gabriel, Chino, and San Bernardino basins in the Los Angeles metropolitan area.
\end{abstract}

\section{Introduction}
Seismic tomography for sedimentary basins is important for assessing earthquake hazards considering the fact that the basins can trap and amplify seismic waves, leading to stronger and longer ground shaking. Fine-scale subsurface structures are typically determined by active source surveys, which often operate at high frequencies but are expensive and limited by environmental impact issues. Alternatively, seismic data from passive sources, such as ambient noise and earthquakes, can be more readily acquired and can also compensate for the absence of low-frequency information in active source data. Many have shown in seismic interferometry that the cross-correlation between diffuse waveforms recorded at two stations approximates the elastodynamic Green's function \citep{wapenaar2004retrieving,wapenaar2010tutorial}. The so-called empirical Green's function (EGF) created from ambient noise is more commonly used by ray-theory-based tomography methods \citep{guo2015high,lin2008surface,yao2010heterogeneity,zheng2011crust}, which do not account for the finite-frequency effects in heterogeneous media. Full waveform inversion (FWI), which accounts for the full physics of wave propagation, can reveal more accurate and detailed subsurface structures \citep{liu20173,liu2023ambient,maguire2022magma,sager2020global,wang2021adjoint,zhang2021rayleigh}. 

Despite its significant advantages, FWI can be time-consuming and memory-intensive for large problems, which hinders its broader applicability. Moreover, the adjoint-state method \citep{fichtner2006adjoint1,liu2006finite,plessix2006review,tromp2005seismic}, which is conventionally used for FWI, requires the initial model to be sufficiently close to the true Earth to avoid being trapped in local minima \citep{gauthier1986two,fichtner2006adjoint2,panning2009seismic,virieux2009overview}. In ambient noise tomography (ANT), the adjoint-state method typically relies on a model derived from surface wave dispersion analysis and ray-based approximations as a starting point \citep{chen2014low,liu2023ambient,maguire2022magma,zhang2018linear,zhang2020wave}. To tackle these challenges, recent studies have shown that neural operators \citep{li2020fourier,li2020neural,azizzadenesheli2024neural} can solve wave equations orders of magnitude faster than traditional numerical solvers \citep{yang2021seismic,yang2023rapid,zou2024deep}. In deep learning, an analogue to the adjoint-state method is known as the reverse-mode automatic differentiation (AD) \citep{baydin2018automatic,rumelhart1986learning}. While these two methods have been shown to be mathematically equivalent \citep{richardson2018seismic,wang2021elastic,Zhu}, AD, which underpins the modern deep learning platforms such as PyTorch \citep{pytorch} and TensorFlow \citep{TensorFlow}, remains aligned with advances in state-of-the-art optimization techniques. For example, methods such as mini-batching \citep{lecun2002efficient}, stochastic gradient descent (SGD) \citep{bottou2010large}, and adaptive moment estimation (Adam) \citep{Kingma} have been developed to better navigate non-convex loss landscapes and escape poor local minima. These methods have prevailed for over a decade in the machine learning community, while remaining relatively underexplored in the classic FWI world. Only in recent years have they begun to attract attention \citep{bernal2021accelerating,mao2025automatic,richardson2018seismic,sun2020theory,thrastarson2022data,van2020accelerated}. Although efforts have been made to implement finite difference solvers in PyTorch/TensorFlow to enable access to AD and various optimization strategies \citep{richardson2018seismic,Zhu}, this approach is memory prohibitive for realistic applications, because back-propagation requires storing wavefields at every time step in memory. We bypass this obstacle by learning a direct mapping through neural operators developed in PyTorch. In alignment with the rapid advancements in optimization techniques, neural operators have the potential to alleviate — though not fully resolve — the cycle-skipping problem in FWI, making it possible to start from a simpler velocity model (e.g., a 1D model).

The primary advantage of neural operators in solving partial differential equations (PDEs) over other prevailing machine learning approaches, such as physics-informed neural networks (PINNs) \citep{raissi2019physics}, is that they are trained to learn the solution operator for an entire family of PDEs instead of a specific instance. This means that a neural operator has the potential to generalize to arbitrary PDE coefficients (e.g., velocity structures) if the training instances are ideally sampled — for example, from random fields \citep{yang2021seismic,yang2023rapid,zou2024deep}. The training is a one-time effort for the forward process and there is no further training for the neural operator in the inversion stage, where the velocity parameters are updated with gradients computed from AD. This is similar to traditional FWI but replaces the adjoint simulation with back-propagation \citep{LeCun}.

In recent years, neural operators have been increasingly applied to seismic wave propagation modeling. They have been used for modeling acoustic \citep{yang2021seismic}, elastic \citep{li2025feature,yang2023rapid}, and viscoelastic \citep{wei2022small} waves. Most applications are limited to 2D due to computational and memory constraints. However, \cite{lehmann20243d,lehmann2025multiple} circumvented the 3D computational bottleneck by predicting only the surface ground motion, while \cite{zou2024deep} modeled the full 3D wavefield by parameterizing the operator in the frequency domain, thereby eliminating the need to model the time dimension. Learning frequency-domain solutions offers memory efficiency for neural operators and has attracted growing interest \citep{cheng2025seismic,huang2025learned,kong2025reducing,li2023solving,zhang2023learning}. Among these, \cite{huang2025learned} and \cite{cheng2025seismic} proposed learning the residual (scattered) wavefield instead of the full wavefield to address the point-source singularity. Current research has primarily focused on synthetic data, while the application of neural operators to real seismic data remains largely unexplored.

Our contributions are as follows. We demonstrate the first application of neural operators for ambient noise tomography on a real seismic dataset, specifically the Basin Amplification Seismic Investigation (BASIN) survey \citep{BASIN2018,clayton2019exposing}, which deployed linear nodal arrays in the northern Los Angeles (LA) basins. We provide a trained neural operator that is applicable without retraining to any linear array in the study area, or potentially other problems with a comparable spatial footprint. With the trained neural operator, FWI can be performed two orders of magnitude faster than the conventional adjoint method. Our tomography results are consistent with previous studies in the same region \citep{ghose2023basin,li2023shear,villa2023three,Zou2024} without the strong dependence on priors. The proposed method can be scaled to 3D, provided that sufficient computing resources are available. 

\section{Methods}
In this section, we first present the governing equations, as well as the methodology of using neural operators to solve the forward and inverse problems. Then, we share the training strategy.
\subsection{Helmholtz Neural Operator with Automatic Differentiation}
\begin{figure}
\centering
\includegraphics[width=1.0\textwidth]{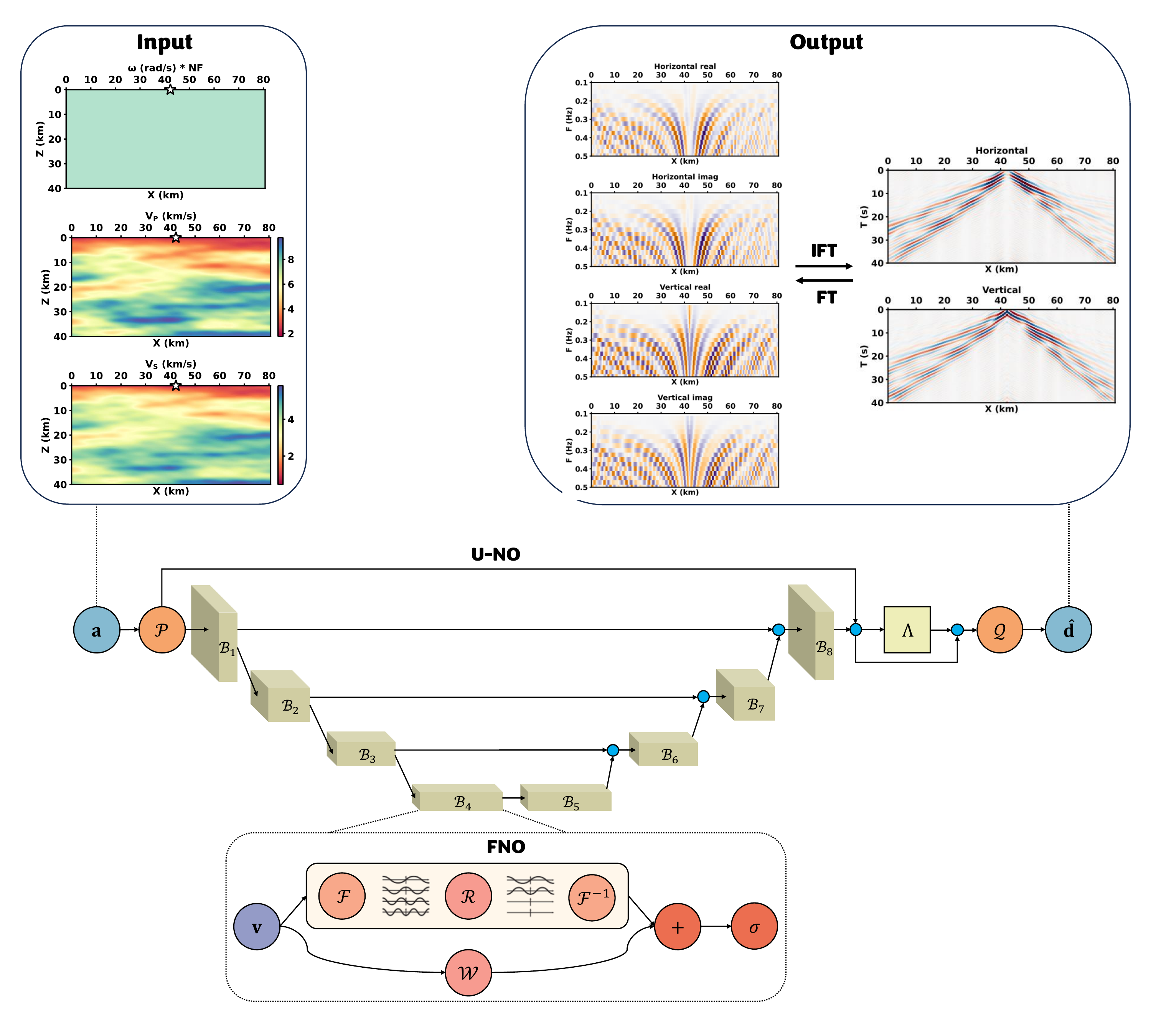}
\caption{Model architecture. $\mathbf{a}$ is the input given by $V_P$, $V_S$, the source location, and a constant function indicating the frequency value. $\mathbf{\hat{d}}$ is the predicted data. $\mathcal{P}$ is a point-wise operator used to lift the dimension. $\mathcal{Q}$ is used to project the output to the desired dimension. $\mathcal{B}$ is the inner integral operator chosen as the FNO, in which $\mathbf{v}$ is input to the layer, $\mathcal{F}$ and $\mathcal{F}^{-1}$ denote the Fourier and inverse Fourier transforms, respectively, $\mathcal{R}$ is a linear operator, $\mathcal{W}$ acts as a residual connection, and $\sigma$ is a nonlinear activation function. $\Lambda$ is a GNO used to query the waveforms at the free surface. Blue circles denote concatenation along the channel dimension.}
\label{HNO}
\end{figure}

Seismic wave propagation in an elastic medium follows the (isotropic) elastic wave equation
\begin{equation}
\begin{aligned} 
 \rho \frac{\partial^2 \mathbf{u}}{\partial t^2}= \nabla \lambda\left ( \nabla \cdot \mathbf{u} \right )+\nabla \mu \cdot \left ( \nabla \mathbf{u}+\left ( \nabla \mathbf{u} \right )^{T} \right )+ \left ( \lambda +2\mu  \right )\nabla\left ( \nabla\cdot \mathbf{u} \right )-\mu \nabla\times \nabla\times \mathbf{u}+\mathbf{f},
\label{ewe}
\end{aligned}
\end{equation}
where $\mathbf{u}$ is the displacement wavefield (three-component vector in 3D media), $t$ is time, $\nabla$ is the gradient with respect to the space coordinates, $\mathbf{f}$ is the body force (source term), $\rho$ is the density, $\lambda$ and $\mu$ are the Lamé parameters. Equation \ref{ewe} can be restated concisely with an integro-differential operator $\mathcal{L}$ as
\begin{equation}
\begin{aligned} 
\mathcal{L}\mathbf{u}=\mathbf{f}.
\label{concise}
\end{aligned}
\end{equation}
In seismology the medium $\mathbf{m}$ is commonly parametrized by P-wave velocity $V_P$ and S-wave velocity $V_S$:
\begin{equation}
V_{P}=\sqrt{\frac{\lambda +2\mu }{\rho }}\quad \text{and} \quad V_{S}=\sqrt{\frac{\mu }{\rho }}.
\label{ps}
\end{equation}
The direct mapping from physical parameters $\mathbf{m}$ to observed waveforms $\mathbf{d}$ is nonlinear and unknown in closed form:
\begin{equation}
    \begin{aligned}
        \mathbf{d} = \mathbf{G}(\mathbf{m}).
        \label{map}
    \end{aligned}
\end{equation}
Based on the universal approximation theorem, neural operators can approximate arbitrary nonlinear continuous operators \citep{hornik1989multilayer,Kovachki}. The solution operator can be parameterized much more efficiently in memory when solving Equation \ref{concise} in the frequency domain (i.e., the Helmholtz equation) than in the time domain \citep{huang2025learned,kong2025reducing,zou2024deep}. This is because different frequency components can be solved individually and, thus, parallelized. In this study, we employ the Helmholtz Neural Operator (HNO) proposed by \cite{zou2024deep}. The HNO computes the solution for each frequency component independently by defining the batch of data at the frequency level. In other words, a batch can contain multiple frequency components due to data shuffling. We use a single HNO to learn the entire frequency band of interest ($0.1$ to $0.5$ Hz), although \cite{kong2025reducing} found that using multiple models to handle different frequencies slightly improves performance. Solutions in the time and frequency domains can be related via the (inverse) Fourier transform. We construct a U-shaped HNO comprising $L=8$ inner layers of Fourier Neural Operators (FNOs) \citep{li2020fourier}, followed by a Graph Neural Operator (GNO) \citep{li2020neural}, to predict the waveforms at the free surface (data, denoted by $\mathbf{\hat{d}}$):
\begin{equation}
\begin{aligned} 
&\mathbf{v}_{0}\left ( x \right )=\left ( \mathcal{P}\,\mathbf{a} \right )\left ( x \right ),\\
&\mathbf{v}_{l+1}\left ( x \right )=\sigma \left ( \mathcal{W}_{l} \,\mathbf{v}_{l}\left ( x \right )+\mathcal{F}^{-1}\left ( \mathcal{F}\left ( \kappa _l \right )\cdot\mathcal{F}\left ( \mathbf{v}_l \right ) \right )\left ( x \right )\right )  \quad \text{where} \quad l=0,...,L-1,\\
&\mathbf{v}_{gno}\left ( x \right )=\int_{B_r(x)} \kappa\left(x, y, \mathbf{v}_{L}\right) \mathbf{v}_{L}(y) \, \mathrm{d}y,\\
&\mathbf{\hat{d}}\left ( x \right )=\left ( \mathcal{Q}\,\mathbf{v}_{gno} \right )\left ( x \right ),
\label{no}
\end{aligned}
\end{equation}
where $\mathbf{a}$ is the input including $V_P$, $V_S$, the source location, and a constant function indicating the frequency value, $\mathbf{\hat{d}}$ is the predicted data for the given frequency component (including the real and imaginary parts), $\mathbf{v}_l$ is input to the ${l+1}^{th}$ FNO layer ($l=0,...,7$) or the GNO ($l=8$), $\mathbf{v}_{gno}$ is the GNO output, $\mathcal{P}$ is a point-wise operator used to lift the dimension, $\mathcal{Q}$ is used to project the output to the desired dimension, $\mathcal{W}_l$ acts as a residual connection, $\kappa_{l}$ is a parametric kernel function, $\mathcal{F}$ and $\mathcal{F}^{-1}$ denote the Fourier and inverse Fourier transforms, respectively, and $\sigma$ is a nonlinear activation function, which we set as GELU \citep{hendrycks2016gaussian}. The GNO uses a kernel function $\kappa$ parameterized as a three-layer neural network, which takes positional information and the output of the last FNO layer as input. $B_r$ denotes the predefined neighborhood of the queried position $x$, defined as all points below the corresponding surface location for computational efficiency. We use FNOs to speed up the integral calculation, and the U-shape allows for a deeper model and skip connections \citep{li2020multipole,Rahman,Ronneberger}. The GNO can improve predictions at queried points and provide greater data efficiency \citep{zou2024deep}. Figure \ref{HNO} illustrates the model architecture and Table S1 provides the details in each neural operator layer. 

In conventional FWI, the solution is obtained by minimizing a scalar-valued objective function $\Phi$, which can, for example, be defined as half the squared $\ell_2$-norm of the data residual:
\begin{equation}
    \begin{aligned}
        \Phi = \frac{1}{2}\left\| \mathbf{\hat{d}} - \mathbf{d}\right\|_{2}^{2} \quad \text{where} \quad \mathbf{\hat{d}}=\mathcal{S}\mathbf{u},
        \label{obj}
    \end{aligned}
\end{equation}
where $\mathbf{\hat{d}}$ is the modeled (predicted) waveforms queried at receivers through a sampling operator $\mathcal{S}$ and $\mathbf{d}$ is the observed data . The parameters of interest $\mathbf{m}$ are updated iteratively using a gradient-based method. Using the adjoint-state method, the gradients of the objective with respect to parameters of interest can be estimated as the cross-correlation between the forward wavefield $\mathbf{u}$ and the adjoint wavefield $\mathbf{u}^*$ at zero time lag, weighted by an analytically derived term \citep{devito}:
\begin{equation}
    \begin{aligned}
        \nabla_\mathbf{m} \Phi = -\mathbf{u}^T \left ( \frac{\partial \mathcal{L}}{\partial \mathbf{m}} \right )^T\mathbf{u}^*.
        \label{adjoint}
    \end{aligned}
\end{equation}
The adjoint wavefield $\mathbf{u}^*$ is generated by back-propagating the residual $\delta \mathbf{d}$ in the medium:
\begin{equation}
    \begin{aligned}
        \mathcal{L}^T \mathbf{u}^*=\mathcal{S}^T\delta \mathbf{d} \quad \text{where} \quad \delta \mathbf{d}=\mathbf{\hat{d}} - \mathbf{d}.
        \label{backprop}
    \end{aligned}
\end{equation}
The adjoint-state method requires the initial model to be sufficiently accurate, with a travel-time error of less than half the period \citep{beydoun1988first,virieux2009overview,Tariq2016}. Otherwise, the cycle-skipping phenomenon may arise, leading to convergence to a local minimum \citep{gauthier1986two,pladys2021cycle}.

With neural operators (or other machine learning methods developed on the PyTorch platform), the gradients of the objective with respect to any parameters in the computational graph are automatically computed through back-propagation \citep{rumelhart1986learning}. This is termed the reverse-mode AD \citep{baydin2018automatic,elliott2018simple}. Provided that all the functions, expressions, and control flow structures used in neural operators are differentiable and compatible with AD, the adjoint simulation can be replaced by back-propagation \citep{LeCun}. It has been shown that the adjoint-state method and reverse-mode AD are mathematically equivalent \citep{richardson2018seismic,wang2021elastic,Zhu}. However, the HNO-AD method eliminates the need to manually derive the gradient in Equation \ref{adjoint} on a case-by-case basis. Moreover, it can take advantage of state-of-the-art optimization techniques built into PyTorch — such as mini-batching and Adam — which help improve the optimization dynamics of FWI and mitigate cycle-skipping. Adam, in particular, a first-order gradient-based optimization algorithm that dynamically rescales the gradients for each parameter based on its past gradients and their squares \citep{Kingma}, has been shown to outperform other optimizers such as SGD and limited-memory Broyden-Fletcher-Goldfarb-Shanno (L-BFGS) \citep{bernal2021accelerating,richardson2018seismic,sun2020theory}.

We summarize the differences between the HNO-AD and adjoint-state methods as follows:
\begin{enumerate}
    \item Forward modeling: In the adjoint-state method, forward modeling is performed by numerically solving the wave equation. In the HNO-AD method, it is performed via a mapping learned by a neural operator.
    \item Gradient computation: In the adjoint-state method, the gradient is analytically derived and computed by running an adjoint simulation. In the HNO-AD method, this is replaced by automatic back-propagation.
\end{enumerate}

\subsection{Data-Driven Training}
In this study, we train a single 2D HNO for linear nodal arrays across the northern LA basins. The model is trained in a supervised learning manner using synthetic data generated from a spectral element method (SEM) using the software SALVUS \citep{Afanasiev}. The computational domain is set to $80$ km (horizontal) $\times$ $40$ km (vertical) on a $256 \times 128$ mesh, which accommodates the longest seismic line, SB1, in the area. We simulate $40$-s-duration wavefields with a time step of $0.001$ s, which are excited by a vertical force and Ricker wavelet time function with a central frequency of $0.3$ Hz. For ambient noise data applications, the source is randomly placed along the free surface from a uniform distribution of horizontal position. The $V_P$ and $V_S$ models are generated from random fields sampled around a background 1D model, which is averaged along the SB1 line from a reference model named CVM-S4.26 \citep{lee2014full}. The parameters for the random fields are provided in Table S2. The density models are derived from $V_P$ using the empirical relation by \cite{brocher2005empirical}, which are input to SALVUS but not to the HNO. The time-domain solutions from the SEM solver are Fourier-transformed and filtered to the frequency band of interest ($0.1$ to $0.5$ Hz) for use with the HNO. We train an HNO using $27000$ simulations and validate the model performance using $3000$ simulations (where a simulation means a single source and a particular velocity model). We define the loss function as a combination of relative $\ell_1$-norm and $\ell_2$-norm of the data residual:
\begin{equation}
\begin{aligned} 
 Loss=0.95\frac{\left\| \mathbf{d}-\mathbf{\hat{d}}\right\|_{1}}{\left\| \mathbf{d}\right\|_{1}}+0.05\frac{\left\| \mathbf{d}-\mathbf{\hat{d}}\right\|_{2}}{\left\| \mathbf{d}\right\|_{2}},
\label{lossfunc}
\end{aligned}
\end{equation}
where $\mathbf{d}$ and $\mathbf{\hat{d}}$ denote the true and predicted data, respectively. We use an Adam optimizer with a learning rate of $0.001$ and a scheduler that decays the learning rate by half every $30$ epochs. The batch size is $256$. Figure S1 shows the loss curves. The full training process took $2$ days with $8$ NVIDIA RTX A6000 GPUs. Note that the training is a one-time effort and no further training is required for different seismic lines or the inversion stage, granted the data have a similar scale. Table \ref{speed} compares the time consumed by SALVUS and the trained HNO in forward modeling and FWI under the same configuration. In performing FWI for this 2D case, the HNO with AD is approximately two orders of magnitude faster than the SEM with the adjoint method. The computational advantage will only be greater in 3D \citep{zou2024deep}.

\begin{table*}
\centering
\caption{Comparison of time consumed by SALVUS and HNO in forward modeling and FWI. The experiments are conducted on a community model called CVM-S4.26 using a single NVIDIA RTX A6000 GPU. The forward modeling time is measured for a single event with one source and $234$ receivers on the free surface. The FWI time is measured for one tomographic iteration for the same event configuration.}
\label{speed}
\begin{tabular}{@{}lll}
\hline
& SALVUS & HNO \\
\hline
Forward modeling & 7.18 s & 0.18 s \\
FWI & 128.51 s & 0.45 s  \\
\hline
\end{tabular}
\end{table*}

\section{Synthetic Tests}
\begin{figure}
\centering
\includegraphics[width=1.0\textwidth]{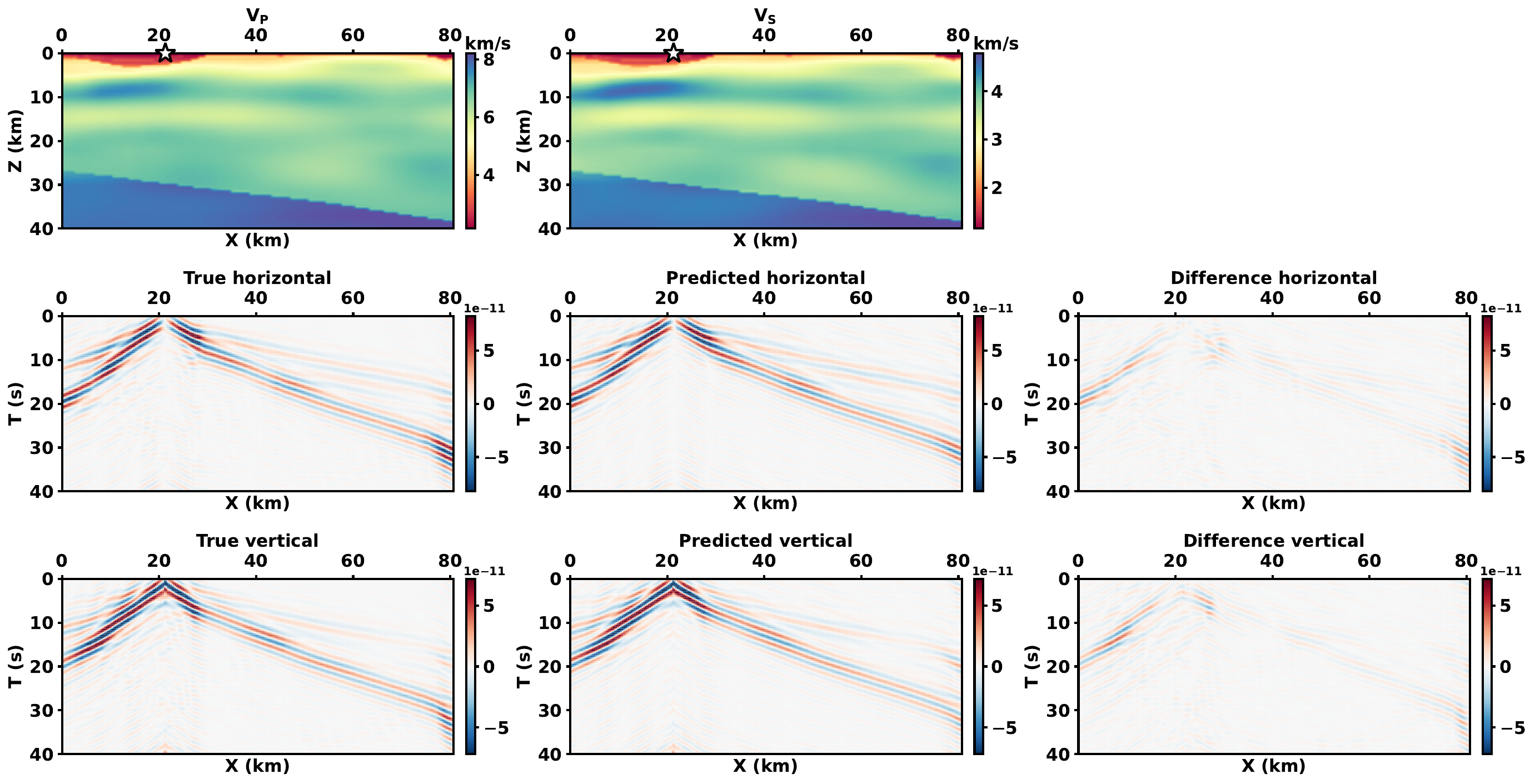}
\caption{Comparing simulations for velocity model CVM-S4.26 between HNO and baseline method. Source location is marked by a white star.}
\label{test_cvms}
\end{figure}

To test generalization, we first evaluate the HNO performance with the synthetic data generated from a community velocity model named CVM-S4.26 \citep{lee2014full}. We extract a profile of the 3D CVM-S4.26 along the longest seismic line, SB1, in the BASIN survey. This model is never explicitly seen by the HNO. Figure \ref{test_cvms} shows the HNO-predicted results for a sample shot excited by a vertical force and Ricker wavelet as in training, with the SALVUS-simulated waveforms serving as the ground truth. The direct output of the HNO is in the frequency domain ($0.1$ to $0.5$ Hz) and is inverse Fourier-transformed to the time domain for visualization purposes. The forward prediction by the HNO is not perfect with a cross-correlation coefficient of 0.98, but we will show that it is sufficiently accurate in the upcoming FWI experiment. 

\begin{figure}
\centering
\includegraphics[width=1.0\textwidth]{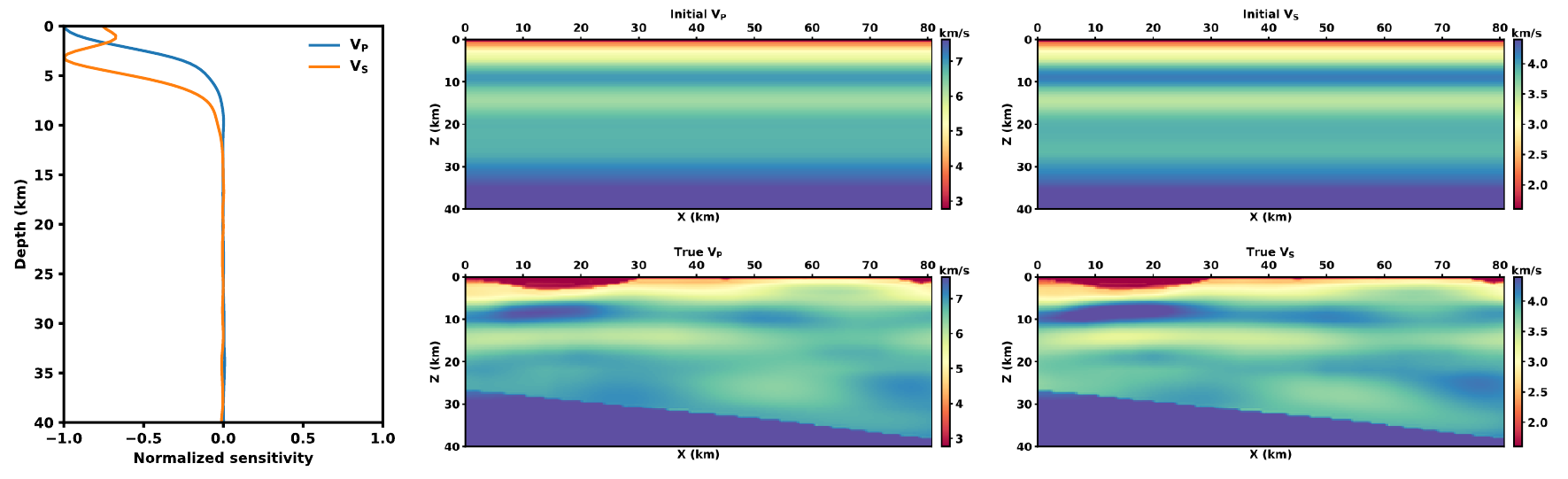}
\caption{Sensitivity kernels for the HNO-AD method, defined as the gradients of the misfit with respect to the velocity parameters, which are summed over all source-receiver pairs and averaged horizontally. The misfit is defined as the squared error between the HNO-predicted data with a 1D initial model and the SALVUS-simulated data with the true model (CVM-S4.26). The sensitivity values for $V_P$ and $V_S$ are each normalized by its maximum amplitude.}
\label{sensitivity}
\end{figure}

Before the inversion, we compute the sensitivity kernels for the HNO-AD method, following the definition in \cite{tromp2005seismic}. In Figure \ref{sensitivity}, we put sources and receivers at nodes on the free surface and sum the gradients of the misfit with respect to the velocity parameters over all source-receiver pairs. The misfit is defined as the squared error between the HNO-predicted data with a 1D initial model and the SALVUS-simulated data with the true model (CVM-S4.26). The 1D initial model is obtained by horizontally averaging the CVM-S4.26 profile. We average the sensitivity kernels horizontally to obtain functions of depth. We can see that the sensitivity drops dramatically with depth, which indicates that updates to deeper parts of the velocity models are less reliable. Therefore, we will only present the inversion results for the top $10$ km hereafter.

\begin{figure}
\centering
\includegraphics[width=0.6\textwidth]{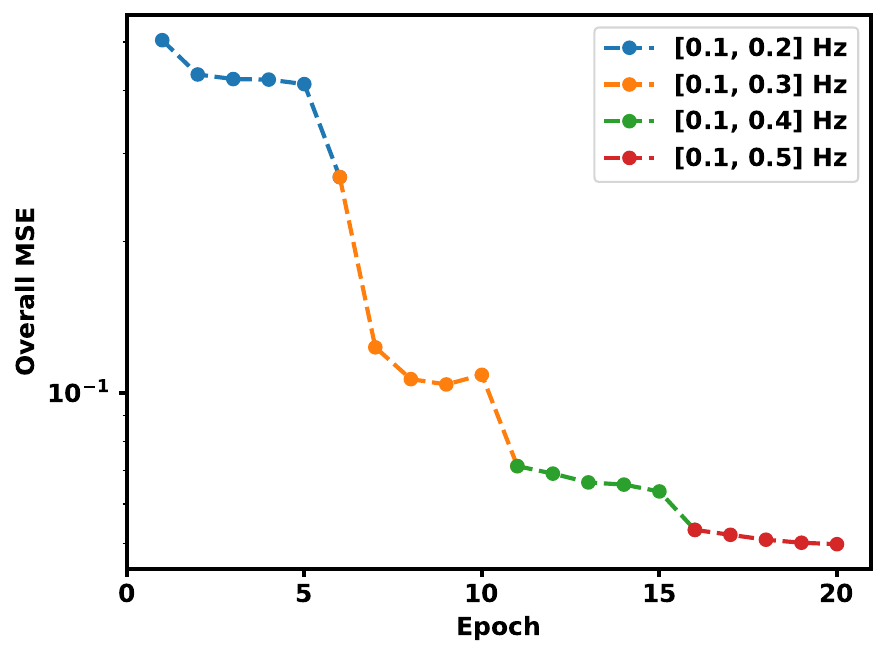}
\caption{Overall MSE (mean squared error) in the FWI process, where higher frequency data is progressively incorporated. We use an Adam optimizer with a learning rate of $0.05$ and a scheduler that decays the learning rate by half every $5$ epochs. The batch size is $64$.}
\label{ltoh}
\end{figure}

In the inversion stage, we freeze the HNO parameters and update $V_P$ and $V_S$ simultaneously with gradients calculated from AD. The inversion is also performed in the frequency domain, and the objective function is defined as the mean squared error (MSE) between the HNO-predicted and SALVUS-simulated data. For use with ambient noise data where the amplitude is often not well-defined, we normalize the amplitude to unity before computing the MSE. This is done by dividing the frequency-domain data, a complex-valued quantity, by its amplitude. We add no regularization term to the objective function but implicitly impose some regularization by smoothing the gradients with a Gaussian filter \citep{tape2010seismic,zhu2015seismic}. The smoothing radius is set to $3\times$ grid spacing ($945$ m) in both horizontal and vertical directions. We average the CVM-S4.26 profile horizontally to get a 1D initial model. We start with lower frequency data and gradually feed in higher frequency components (from $0.1$ to $0.5$ Hz), which is also common practice in traditional FWI to mitigate cycle-skipping \citep{wang2018refined,zhang2018linear}. This is because skipping one cycle in lower frequency data requires larger velocity variation, whereas velocity variation is smaller at greater depth — where lower frequency data has better sensitivity. Also, the neural operator performs more accurately for lower frequencies \citep{zou2024deep}. Figure \ref{ltoh} shows the MSE for the full frequency band in the inversion process. We use an Adam optimizer with a learning rate of $0.05$ and a scheduler that decays the learning rate by half every $5$ epochs. For simplicity, we use a fixed number of epochs for each frequency band, balancing accuracy and efficiency, although adapting the epoch count based on convergence may be a better option. The gradients are summed over a mini-batch of $64$ data points, where a single data point corresponds to a single frequency and a single source. 

\begin{figure}
    \centering
    \includegraphics[width=0.6\textwidth]{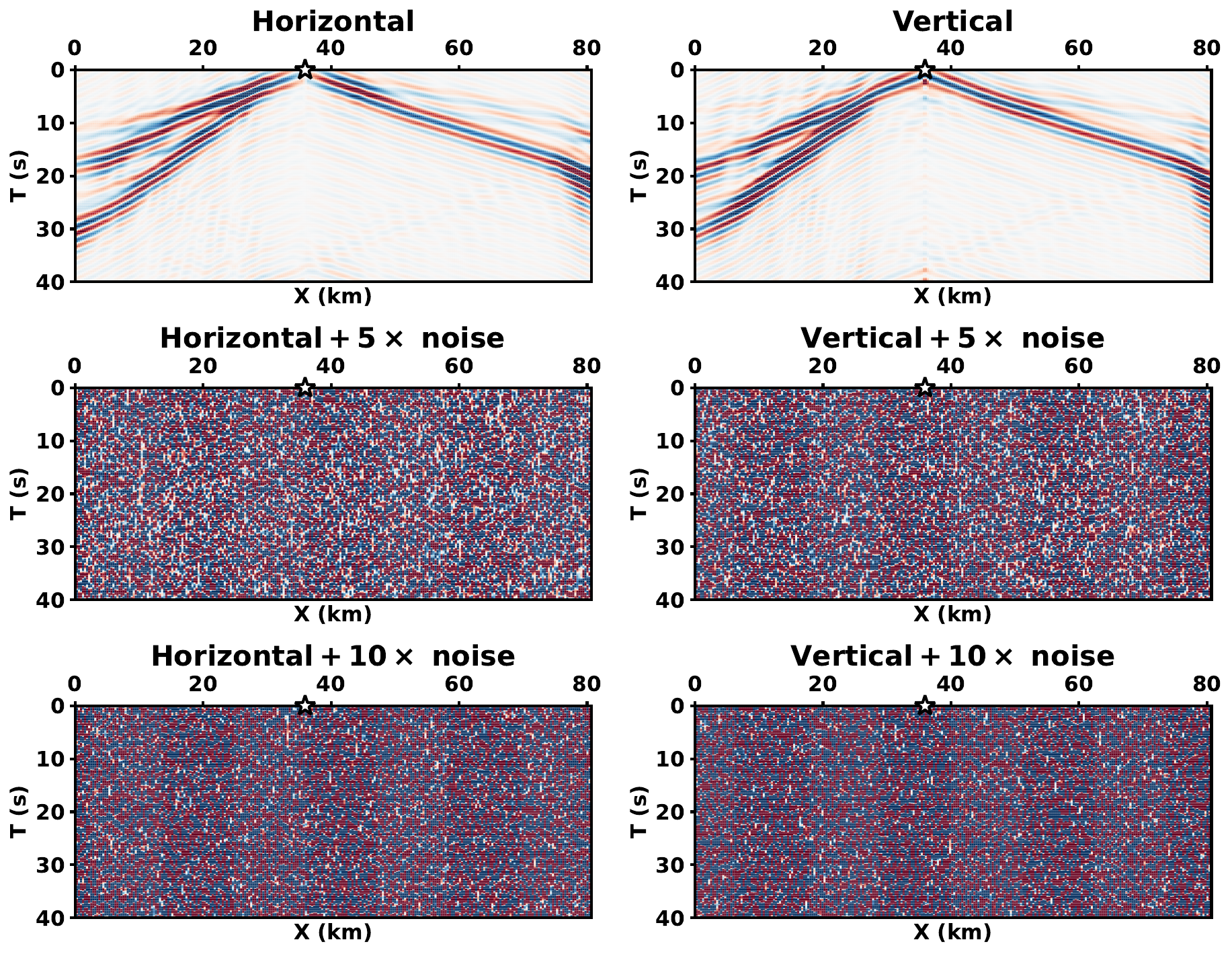}
    \caption{A sample shot of synthetic data perturbed with different levels of noise. The noise is generated from a zero-mean Gaussian distribution with an SD equal to a factor times the SD of the noise-free data.}
    \label{synnoise}
\end{figure}

\begin{figure}  
    \centering
    \includegraphics[width=0.8\textwidth]{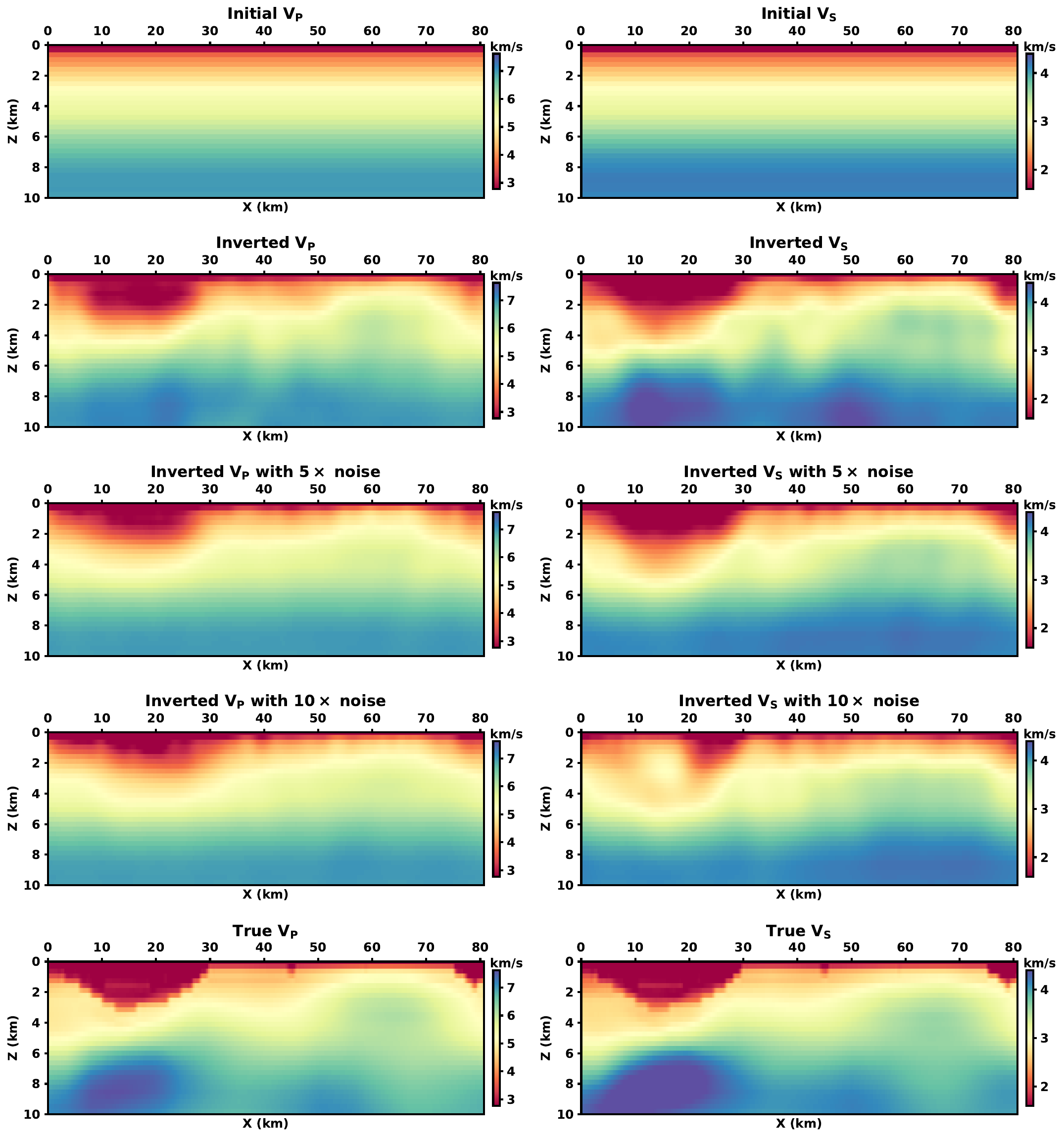}
    \caption{Synthetic FWI results for noise-free data and noisy data.}
    \label{synnoisefwi}
\end{figure}

We conduct FWI experiments on both noise-free data and data perturbed with different levels of Gaussian random noise. We activate $234$ sources along the free surface of the CVM-S4.26 slice for SB1, located at the nodes. The receivers are also placed at the nodes. Figure \ref{synnoise} shows a sample shot of synthetic data perturbed with different levels of noise. The noise is generated from a zero-mean Gaussian distribution with a standard deviation (SD) equal to a factor times the SD of the noise-free data. The corresponding FWI results for noise-free data and noisy data are displayed in Figure \ref{synnoisefwi}. The noise-free inversion in the second row clearly reveals the basin model, demonstrating the forward modeling accuracy of the HNO from another perspective. The deeper part of the model is not well resolved because the source-receiver geometry and frequency content limit sensitivity to near-surface regions. When $5\times$ noise is added, the FWI results maintain good quality, even though the seismic signal already becomes unrecognizable to the eye. This could be explained by the fact that FNOs are global operators, and the zero-mean noise might be partially averaged out in the kernel. However, real noise typically does not behave this nicely. We increase the noise level until the inversion is on the verge of collapse, with a perturbation of $10\times$ noise that totally obliterates the waveforms. 

These synthetic tests serve to verify the generalizability of the neural operator trained with random fields and noise-free data to realistic velocity structures and noisy data, supervised by some definition of ground truth that is unavailable in real data. They show the excellent performance of the HNO-AD method in low signal-to-noise ratio (SNR) situations. Additionally, the entire FWI process can be completed in minutes. In contrast, FWI with the adjoint-state method takes $16$ hours for $10$ iterations and makes little progress with the same 1D initialization even for noise-free data (Figure S2). We attribute the main difference to the use of mini-batch Adam optimization in the HNO-AD method, which is not yet supported by SALVUS. We provide another generalization test in Figure S3 for a velocity model obtained from a ray-theory-based method \citep{li2023shear}, which is fully independent of the training process.

\section{Application to Real Seismic Data}
\begin{figure}
\centering
\includegraphics[width=1.0\textwidth]{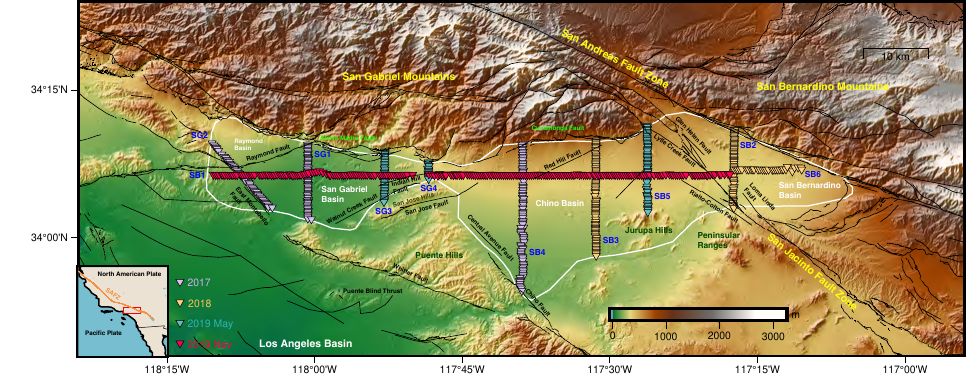}
\caption{Map of the northern Los Angeles basins, adapted from \protect\cite{Zou2024}. The study region is outlined by the white polygon. Triangles in different colors show nodal lines deployed in different time periods.}
\label{area}
\end{figure}

We apply the trained neural operator to a real-world ambient noise tomography example. The real data comes from the BASIN survey that deployed $10$ linear nodal arrays in the San Gabriel (SG), Chino, and San Bernardino (SB) basins north of LA \citep{clayton2019exposing}. A total of $758$ Fairfield ZLand three-component nodes with a corner frequency of $5$ Hz recorded ambient noise for approximately a month. Each line consisted of $14$ to $260$ geophones, spaced approximately $0.25$ km apart. The geometry of the survey was designed to detect subsurface variations up to $0.5$ km in horizontal scale. Figure \ref{area} shows a map of the study area and station configuration. The sedimentary formations date from the opening of the LA basin in the Miocene \citep{wright1991structural}. Details in the geologic and tectonic setting can be found in \cite{villa2023three} and \cite{ghose2023basin}.

We take the EGF derived from ambient noise cross-correlation as the direct observation:
\begin{equation}
\begin{aligned} 
G_{AB}(t)=-\frac{\mathrm{d}}{\mathrm{d} t}\left (\frac{C_{AB}(t)+C_{AB}(-t)}{2}\right ) \quad \text{where} \quad t>0,
\label{egf}
\end{aligned}
\end{equation}
where $G_{AB}(t)$ is the displacement EGF between two stations $A$ and $B$, and $C_{AB}(t)$ is the cross-correlation between the seismograms recorded at these two stations. The positive and negative lags of $C_{AB}$ are averaged to enhance the SNR and to mitigate the effect of inhomogeneous source distribution \citep{lin2008surface}. Although more effective methods for addressing non-diffuse noise fields have been explored \citep{liu2020finite,liu2023ambient}, this is not the focus of this study, and we follow the standard procedure. Our cross-correlation workflow is based on \cite{bensen2007processing} and \cite{li2023shear}. We correlate both daytime and nighttime recordings from the one-month deployment, since using only nighttime data merely reduced the anthropogenic noise in a minor way \citep{li2023shear}. One-hour data segments are correlated and stacked to derive the final correlation. We convert the three-component data from the vertical-north-east (ZNE) to the vertical-radial-transverse (ZRT) coordinate system. The particle motion of Rayleigh waves lies in the ZR plane, whereas Love waves involve mainly horizontal motion in the transverse direction. We demean the raw seismograms and apply a bandpass filter between $0.1$ and $2$ Hz. Prior to the correlation, we perform temporal normalization to suppress the effects of earthquakes and apply spectral whitening to broaden the effective bandwidth.

The EGF obtained from cross-correlation accounts for wave physics in a 3D medium, to which a phase shift of $\pi /4$ must be applied prior to a 2D inversion \citep{forbriger2014line,schafer2014line,zhang2018linear}. Figure S4 shows the effect of this transformation. A factor for amplitude correction related to geometrical spreading is further needed \citep{zhang2018linear}, but we focus solely on the phase information, as the amplitude recovered in ANT is not well-defined. We have verified through synthetic tests that the phase information is sufficient to reconstruct the velocity models (Figure \ref{synnoisefwi}). We eliminate the effect of amplitude differences between 2D and 3D by normalizing the amplitude to unity during inversion. This also helps to bridge the gap between the synthetic Green's function (SGF) and EGF, as they have different source bandwidths. We match the Z-Z and Z-R cross-correlations with the vertical and horizontal waveforms predicted by the HNO, respectively. We mainly retrieve Rayleigh waves with particle motion confined in a 2D plane. In this study, we follow the principle of Green's function retrieval, while \cite{tromp2010noise}, \cite{sager2018towards,sager2020global}, and \cite{tsai2024towards} established the full waveform ambient noise inversion technique that goes beyond that by interpreting correlation functions as self-consistent observables.

We maintain most settings from the previous synthetic FWI results in carrying out the real data experiments, except we include an additional regularization term that encourages the inverted $V_P$ and $V_S$ to respect \cite{brocher2005empirical}:
\begin{equation}
\begin{aligned} 
 L_{reg} = \frac{1}{M}\sum_{i=1}^{M} \left ( V_{P_i} - (0.9409 + 2.0947V_{S_i} - 0.8206V_{S_i}^2 + 0.2683V_{S_i}^3 - 0.0251V_{S_i}^4) \right )^2,
\label{psreg}
\end{aligned}
\end{equation}
where $M$ is the number of grid points on which the velocity models are discretized. This is to constrain $V_P$, which is not well constrained by surface waves.

\begin{figure}
\centering
\includegraphics[width=0.8\textwidth]{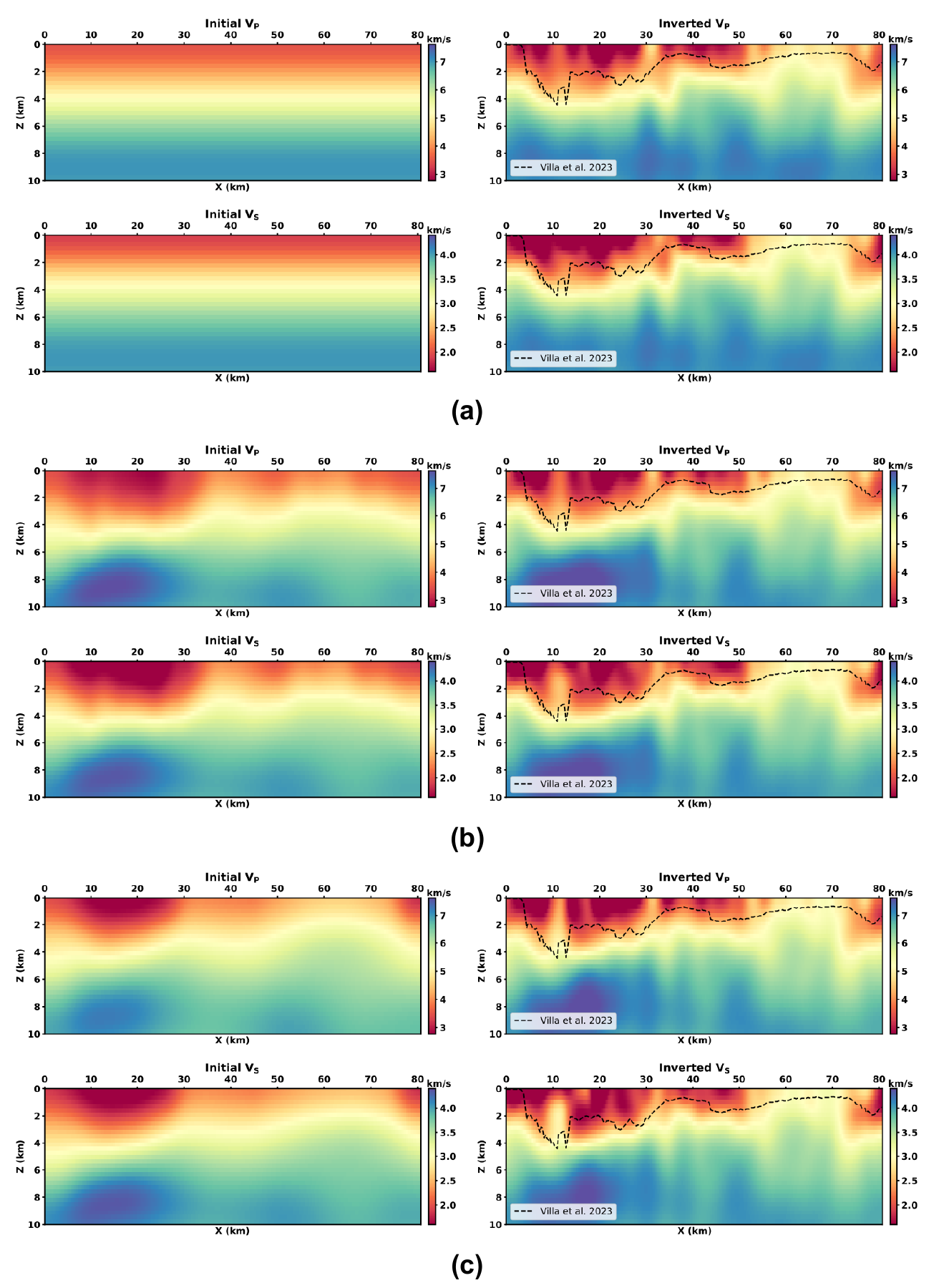}
\caption{Inversion results for SB1 with different initial models including the (a) 1D and (b) smoothed models from \protect\cite{li2023shear}, and (c) the smoothed CVM-S4.26. The along-profile distance increases from west to east. The basin bottom from \protect\cite{villa2023three} is delineated with black dashed lines for reference.}
\label{real}
\end{figure}

We begin the real data FWI with the longest line, SB1, which extends about $80$ km from west to east. Beneath SB1 lie two significant basins — the deeper SG basin in the west and the shallower Chino basin in the east. The depth of the sediment-basement interface has been constrained by \cite{villa2023three}. In addition to this 3D basin depth model and the CVM-S4.26 \citep{lee2014full}, another reference is from \cite{li2023shear} who constructed a 3D shear wave velocity model for the same area through surface wave dispersion analysis. We start the inversion with a 1D model calculated from this reference model, which results in Figure \ref{real}a. Overall, the inverted velocity models reveal basin structures that align well with \cite{villa2023three}, capturing significant features including the deeper and more heterogeneous SG basin in the west and the low-velocity zone at the edge of the relatively flat Chino basin adjacent to the SB basin. We demonstrate the robustness of this result through another two initial models, the smoothed models from \cite{li2023shear} (Figure \ref{real}b) and CVM-S4.26 (Figure \ref{real}c). Results from these different initial models show excellent overall consistency, although they differ in details. In particular, the 1D initial model can achieve performance comparable to that of models with more detailed priors. The corresponding error curves are provided in Figure S5.

\begin{figure}
\centering
\includegraphics[width=1.0\textwidth]{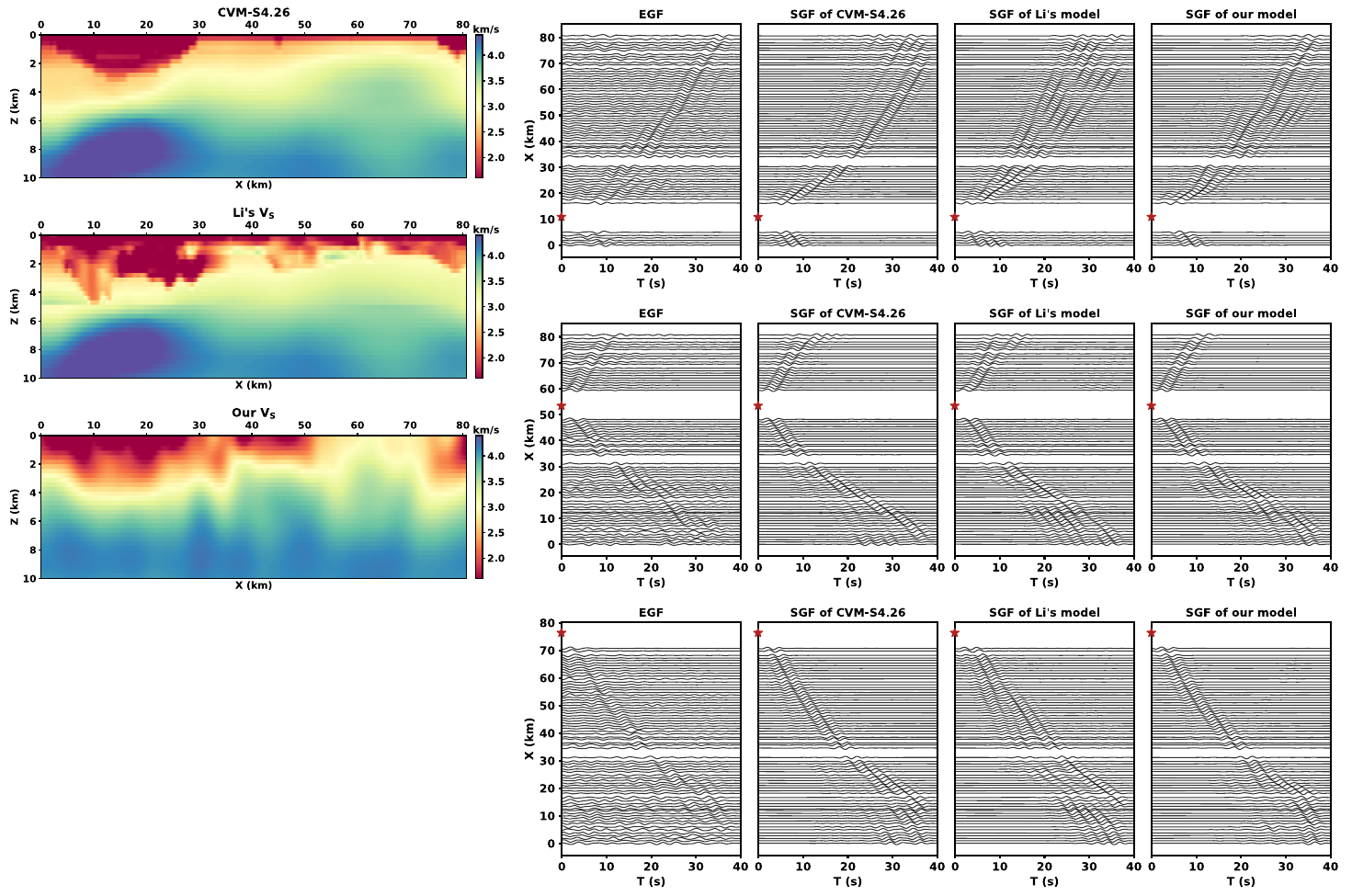}
\caption{Examples of SALVUS-simulated waveforms (SGFs) using the CVM-S4.26, model from \protect\cite{li2023shear}, and our model from Figure \ref{real}a compared with the EGF for SB1. Only $V_S$ is plotted in the top left for simplicity. The frequency band of the data is $0.2$ to $0.5$ Hz. In plotting, the waveform is normalized by trace and every third trace is displayed.}
\label{waveform}
\end{figure}

The incorporation of higher frequency data in later stages of FWI contributes to refining the shallower structure but also introduces high-frequency noise. Some of the noise can potentially be suppressed using a signal window. Meanwhile, the inaccuracy of the neural operator in forward modeling introduces another source of uncertainty. We compare SALVUS-simulated waveforms (SGFs) using the CVM-S4.26, model from \cite{li2023shear}, and our model from Figure \ref{real}a with the EGF for SB1 in Figure \ref{waveform}. The corresponding waveform difference is plotted in Figure S6. The point here is not to conclude that our model is better than others, but rather to demonstrate that the HNO-AD method can achieve accuracy comparable to traditional methods at a much lower cost in terms of time and effort. Also note that our model starts with a 1D initial model, while the other two models have stronger priors \citep{lee2014full,li2023shear}.

\begin{figure}
    \centering
    \includegraphics[width=0.8\textwidth]{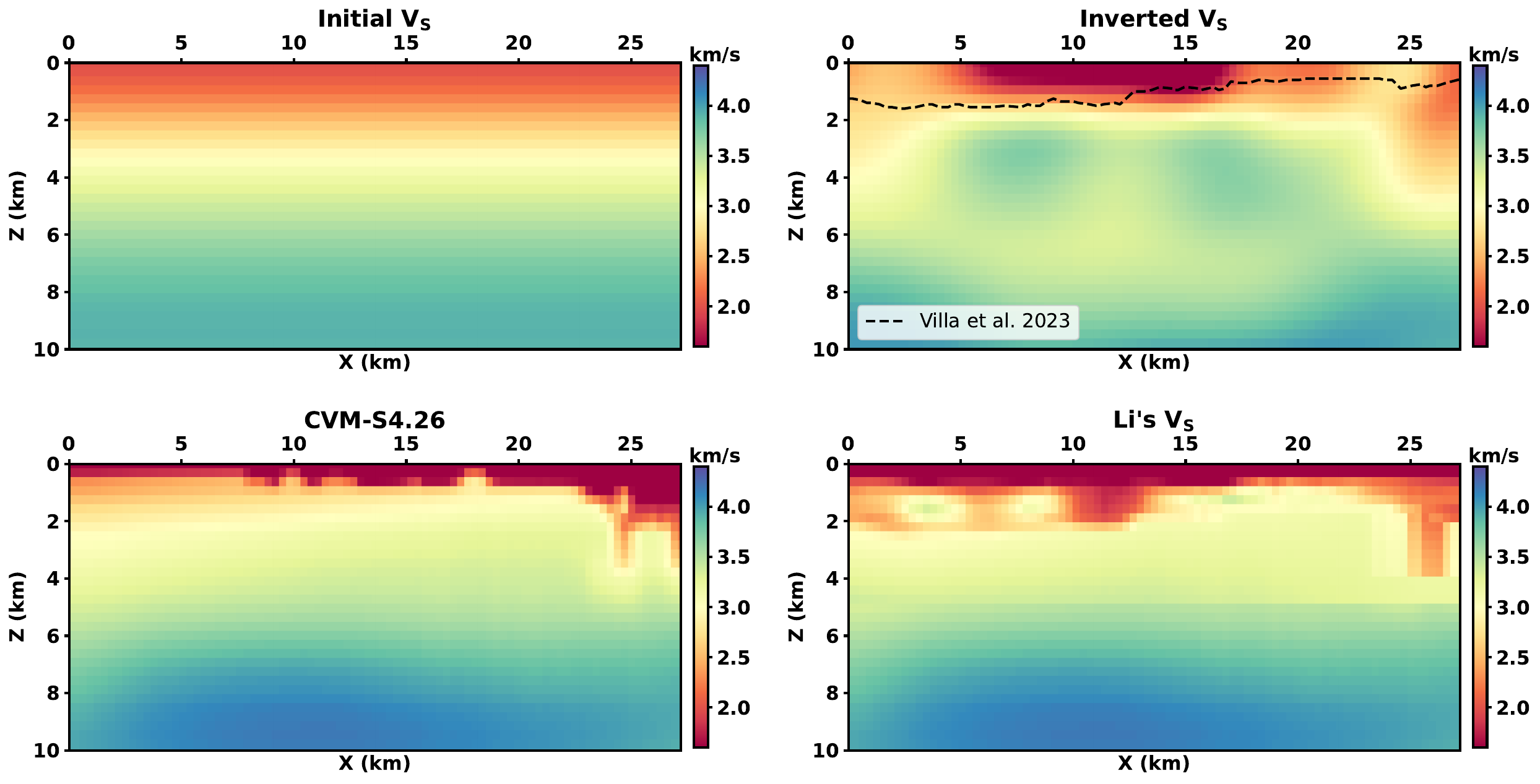}
    \caption{Inversion results for SB4, in comparison to the CVM-S4.26 and model from \protect\cite{li2023shear}. Only $V_S$ is plotted for simplicity. The along-profile distance increases from north to south.  The basin bottom from \protect\cite{villa2023three} is delineated with black dashed lines for reference.}
    \label{SB4inv}
\end{figure}

\begin{figure}  
    \centering
    \includegraphics[width=0.8\textwidth]{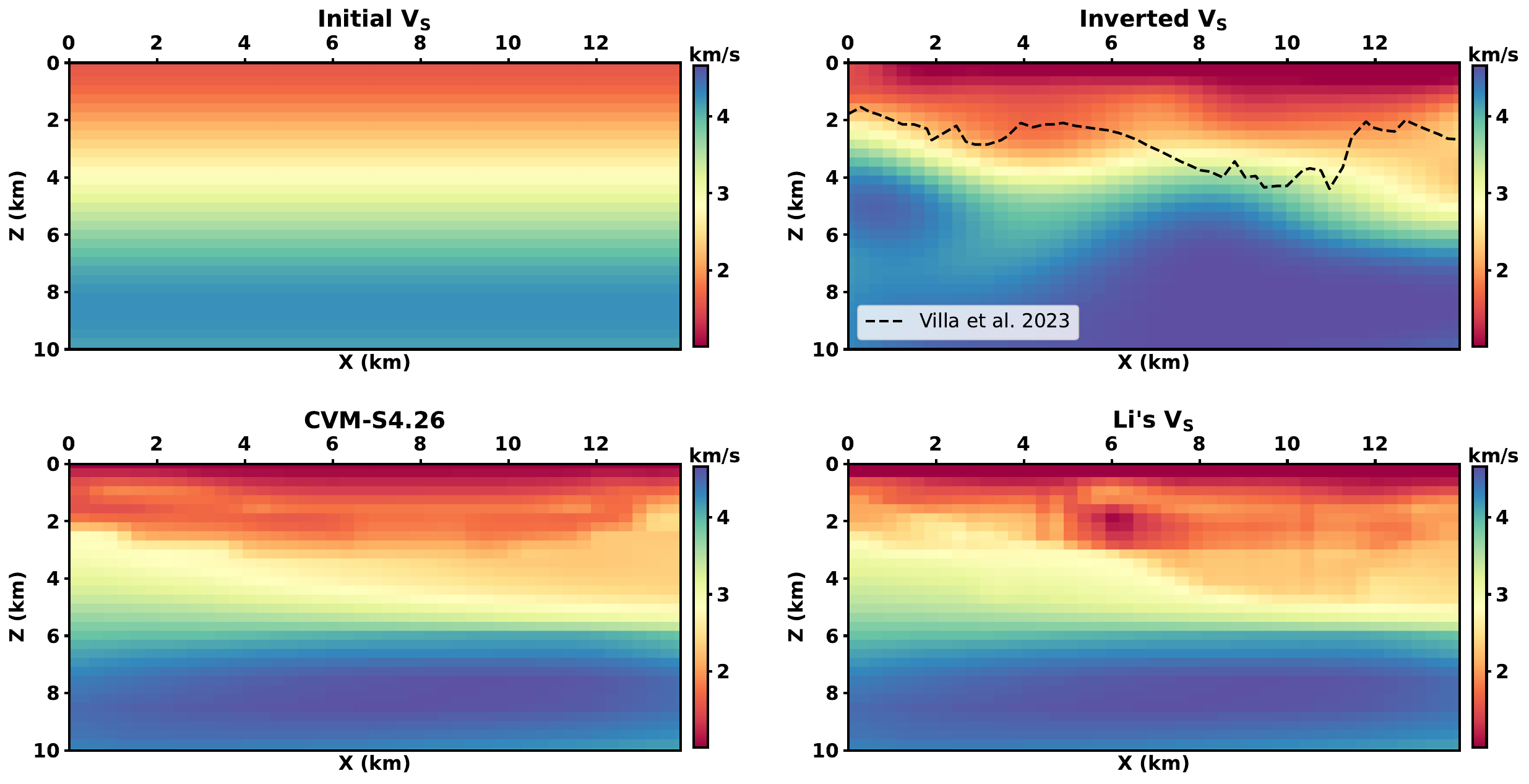}
    \caption{Inversion results for SG1, in comparison to the CVM-S4.26 and model from \protect\cite{li2023shear}. Only $V_S$ is plotted for simplicity. The along-profile distance increases from north to south. The basin bottom from \protect\cite{villa2023three} is delineated with black dashed lines for reference.}
    \label{SG1inv}
\end{figure}

The same neural operator can be directly applied to any seismic line in the study region, because it was trained to cover the longest one. While the model cannot extrapolate to larger domains than it was trained on, application to smaller domains is straightforward by simply extracting a subset of the inversion result. Additionally, no specific orientation is assumed for the 2D profiles, since the training data are generated in a Cartesian coordinate system without reference to geological orientation. This ensures that the model is not biased toward any particular direction. We show another two examples for SB4 and SG1 in Figures \ref{SB4inv} and \ref{SG1inv}, respectively. For SB4 that crossed the Chino basin from north to south, our inversion result reveals a low-velocity zone at the southern end near the Chino fault (Figure \ref{area}), which is also seen in the CVM-S4.26 and \cite{li2023shear}. The low-velocity anomalies between $0$ and $15$ km present in \cite{li2023shear} are absent in both our model and CVM-S4.26. While the authors did not provide specific interpretation, the seismic line crossed the Red Hill fault at approximately $5$ km, which could explain the observed discontinuity (if it is not an artifact). Interestingly, our model shows a velocity jump near the Red Hill fault. Overall, the tomography for SB4 displays a shallow (approximately $2$ km) sedimentary basin in agreement with previous studies. The sediment-basement interface beneath SG1 is much deeper, as captured in all the displayed models. Our result for SG1 aligns with that for SB1 in the SG basin in terms of basin depth but captures less heterogeneity, which might be limited by the array length. These inversions can be completed in a few minutes. 

\section{Discussion}
\begin{figure}
\centering
\includegraphics[width=0.8\textwidth]{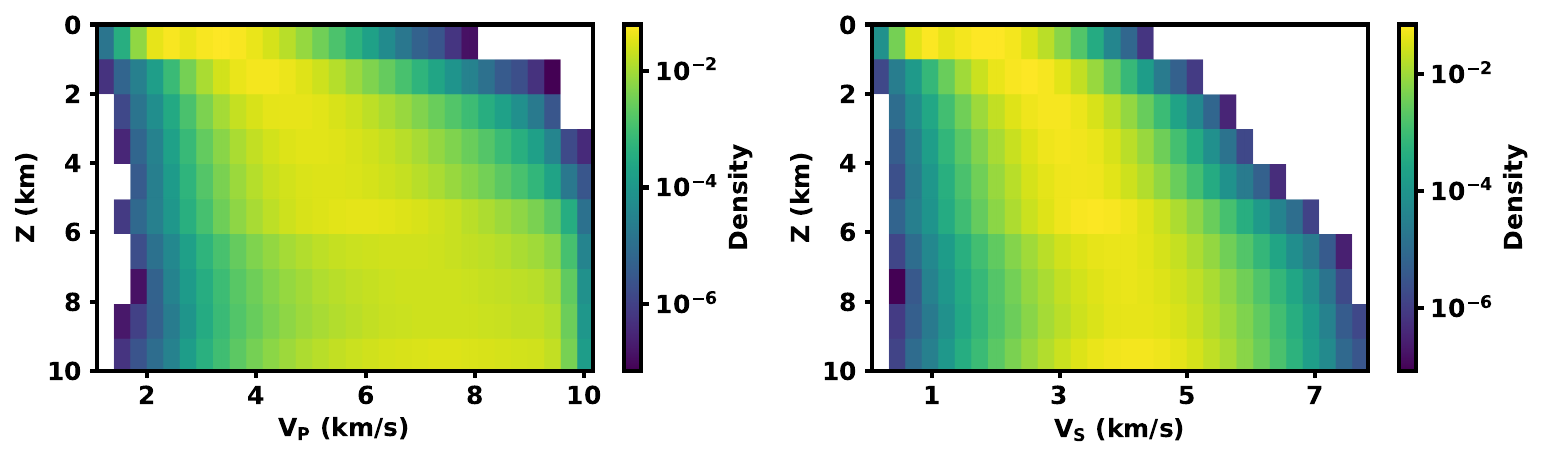}
\caption{Training distributions of $V_P$ and $V_S$ as a function of depth.}
\label{traindist}
\end{figure}

In this study, we have demonstrated the functionality of neural operators in realistic ambient noise full waveform inversion through a 2D case. Provided sufficient computational resources, the proposed method is expected to work in 3D without fundamental changes, while providing increased computational advantages over conventional methods. Because our model is trained within the framework of supervised learning, the fundamental assumption is that training and testing data follow the same distribution. While this assumption seldom holds strictly in practice, it serves as a reference for identifying when model predictions are likely to fail — for example, in cases where the velocity parameters deviate significantly from the training distribution. We show the training distributions of $V_P$ and $V_S$ as a function of depth in Figure \ref{traindist}, allowing users to decide whether to directly use our model or perform additional training. Even in cases where re-training is necessary, a pre-trained model enables transfer learning that can greatly improve learning efficiency \citep{yosinski2014transferable}. Another aspect is that the model is not expected to extrapolate to larger physical domains than which it was trained on ($80$ km length by $40$ km depth) or to frequencies outside the preset band of interest.

In performing FWI, we use gradient descent with AD instead of training a reverse model that directly maps from observed waveforms to velocity parameters. Although such end-to-end approaches are even faster \citep{kazei2021mapping,moseley2020deep,wu2019inversionnet}, they provide only a single solution without supervision on the data misfit. Such approaches are arguably not ideal for inverse problems because they are ill-posed and non-unique, unless they are combined with uncertainty quantification methods. We define the misfit function for inversion as the MSE in the frequency domain for convenient use with HNO, but it would be feasible and worthwhile to explore different misfit functions in either the frequency or time domain \citep{bozdaug2011misfit,metivier2018optimal,fournier2019inversion,sambridge2022geophysical,sun2019application} to better tackle the non-uniqueness and cycle-skipping problems. It is important to note, however, that the AD method requires a differentiable loss function, or some soft approximation to address discontinuity \citep{liu2023robust}.

The rapid developments in compute capabilities and machine learning research have positioned data-driven approaches as a promising new perspective for FWI. On one hand, data-driven approaches can greatly accelerate modeling and inversion. On the other hand, they rely on data provided by physics-driven approaches and, therefore, will also benefit from any advancements in those approaches. In this study, we integrate these two types of approaches by first generating training data with a physics-driven solver and then training a data-driven model as a surrogate solver. Another way is to incorporate physics into the training loss, making it possible to remove the external solver \citep{li2024physics,raissi2019physics,Rasht‐Behesht,ren2024seismicnet,SongAlkhalifah}. While training with physics loss alone is challenging, combining it with a data loss has the potential to improve generalization ability and further benefit inversion.

\section{Conclusions}
We present the first application of neural operators in ambient noise full waveform inversion using several real seismic datasets from linear nodal arrays distributed across the northern LA basins. We show that a Helmholtz Neural Operator trained on random velocity fields can generalize to realistic velocity structures, while demonstrating robustness to random noise. In performing FWI with automatic differentiation, the neural operator is two orders of magnitude faster than the conventional spectral element method and eliminates the need for manual gradient derivation. The HNO-AD method, built on PyTorch, offers enhanced flexibility to escape poor local minima. The tomography results from real data align with multiple previous studies and can be obtained with minimal additional effort. We provide a trained model that can be directly applied to different regions, as long as the velocity parameters fall within the training distribution.

\section*{Data Availability Statement}
The node data can be accessed from the IRIS Data Management Center \citep{BASIN2018}. The network codes are 4M and 6J. The code, models, and processed data are available at \url{https://github.com/caifeng-zou/ANFWI_HNO} \citep{caifeng_zou_2025_15061657}. 

\section*{Acknowledgements}
ZER acknowledges financial support from the David and Lucile Packard Foundation. CZ and RWC acknowledge support from NSF EAR-2105358 and EAR-2438773. FCL acknowledges financial support from the Statewide California Earthquake Center (SCEC) based on Award Number DE-SC0016520 from the U.S. Department of Energy and NSF EAR-2438772. We sincerely thank Fenglin Niu as the editor, Tariq Alkhalifah, and an anonymous reviewer for their careful reading and constructive feedback, which led to substantial improvements in this paper. We thank all the volunteers who helped deploy the nodal arrays and Patricia Persaud for coordinating the BASIN project.

\bibliography{main.bib}
\bibliographystyle{myst}

\input{supple}

\label{lastpage}

\end{document}

%% file: supple.tex
\allowdisplaybreaks

\setcounter{figure}{0} 
\setcounter{table}{0} 
\setcounter{page}{0}

\renewcommand{\thepage}{S\arabic{page}} 
\renewcommand{\thefigure}{S\arabic{figure}}
\renewcommand{\thetable}{S\arabic{table}}
\renewcommand{\thesection}{S\arabic{section}} 
\renewcommand{\theequation}{S.\arabic{equation}} 

\newpage
\section*{Supplementary Materials}

\begin{figure}[htbp]
\centering
\includegraphics[width=0.6\textwidth]{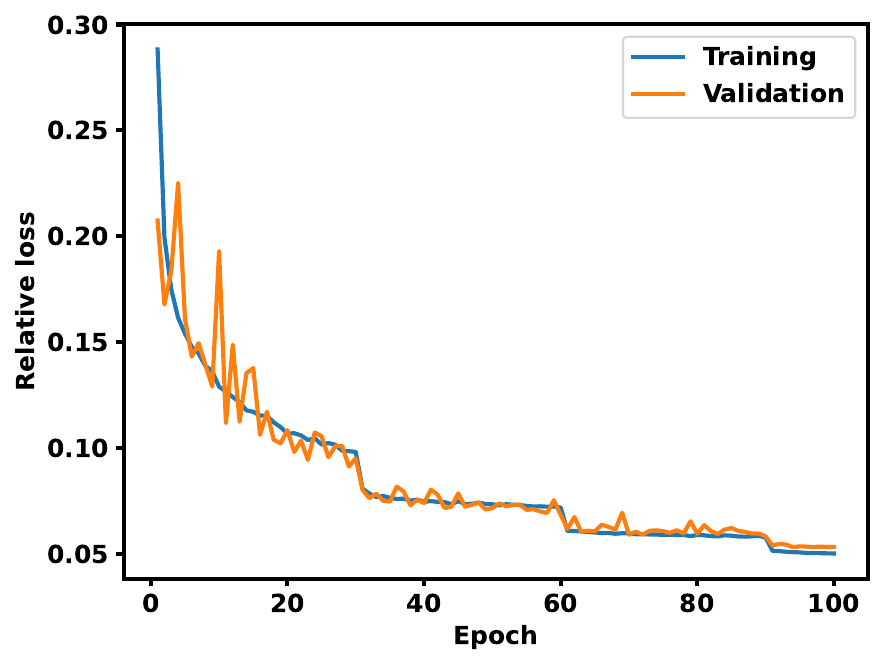}
\caption{Loss curves. The batch size is $256$. We use an Adam optimizer with a learning rate of $0.001$ and a scheduler that decays the learning rate by half every $30$ epochs.}
\label{loss}
\end{figure}

\begin{figure}
\centering
\includegraphics[width=0.9\textwidth]{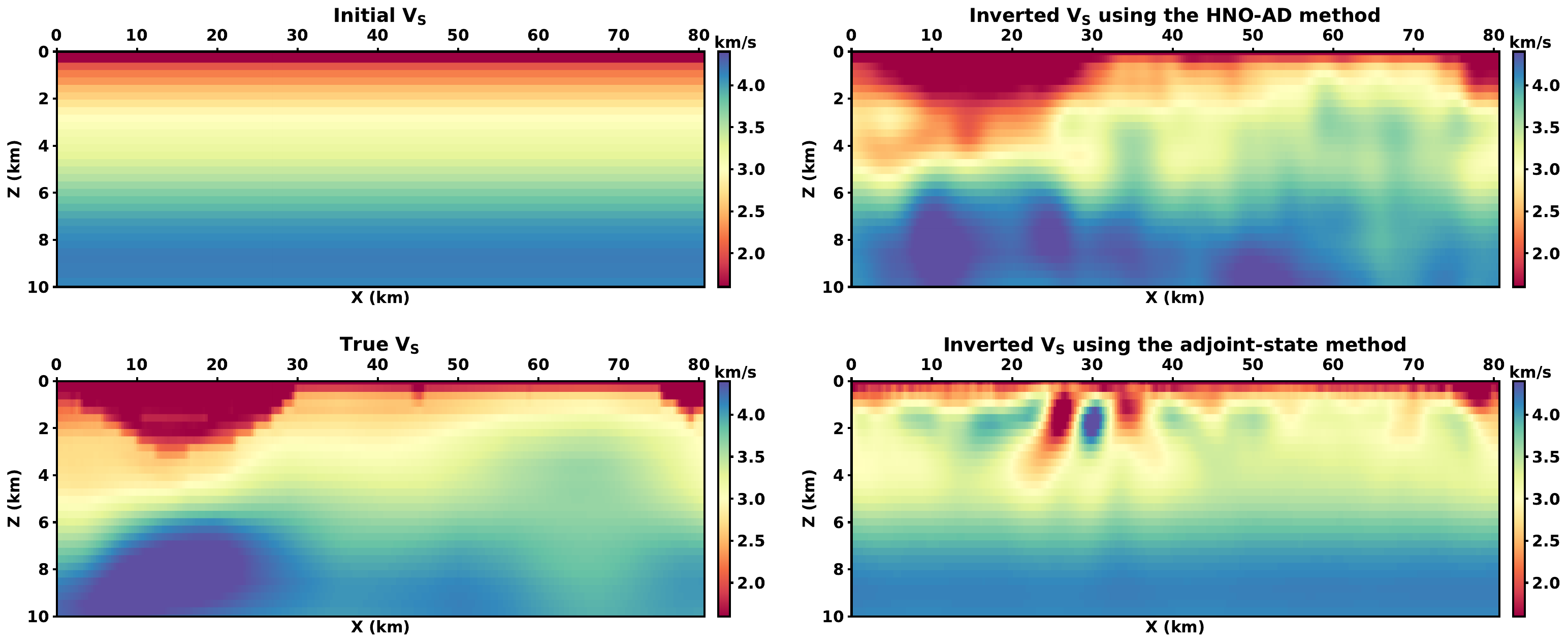}
\caption{FWI results for noise-free data using the HNO-AD and adjoint-state methods with the same initialization. The HNO-AD method uses mini-batch Adam (not yet supported by SALVUS), while the adjoint-state method uses full-batch L-BFGS. The result from HNO-AD is noisier than that in Figure \ref{synnoisefwi}, because all frequencies are handled together here.}
\label{adjointFWI}
\end{figure}

\begin{figure}
\centering
\includegraphics[width=0.8\textwidth]{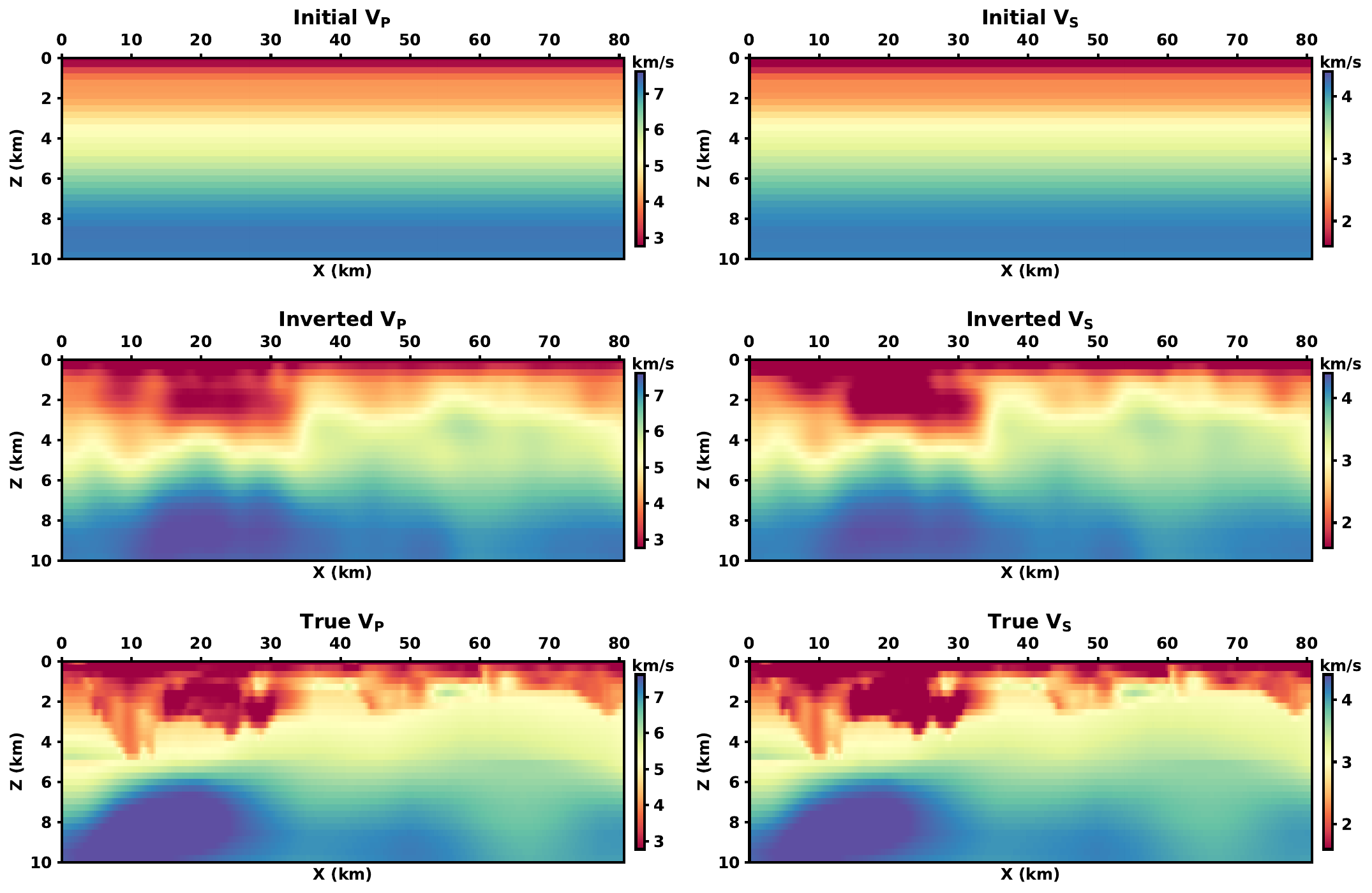}
\caption{Synthetic FWI results for the velocity model from \cite{li2023shear}.}
\label{inv_li_r2}
\end{figure}

\begin{figure}
\centering
\includegraphics[width=0.6\textwidth]{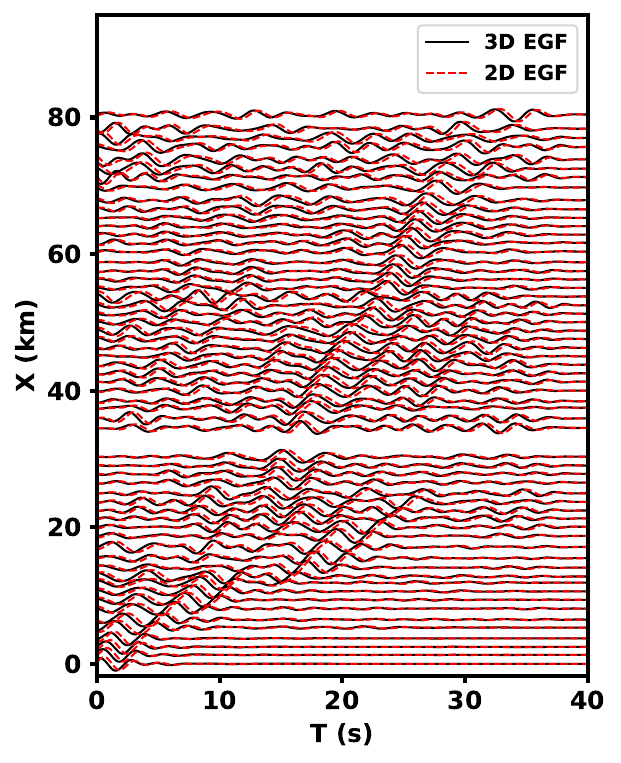}
\caption{3-to-2D converted EGF created from ambient noise cross-correlation for SB1. Every fourth trace is displayed.}
\label{3d_to_2d_egf}
\end{figure}

\begin{figure}
\centering
\includegraphics[width=0.6\textwidth]{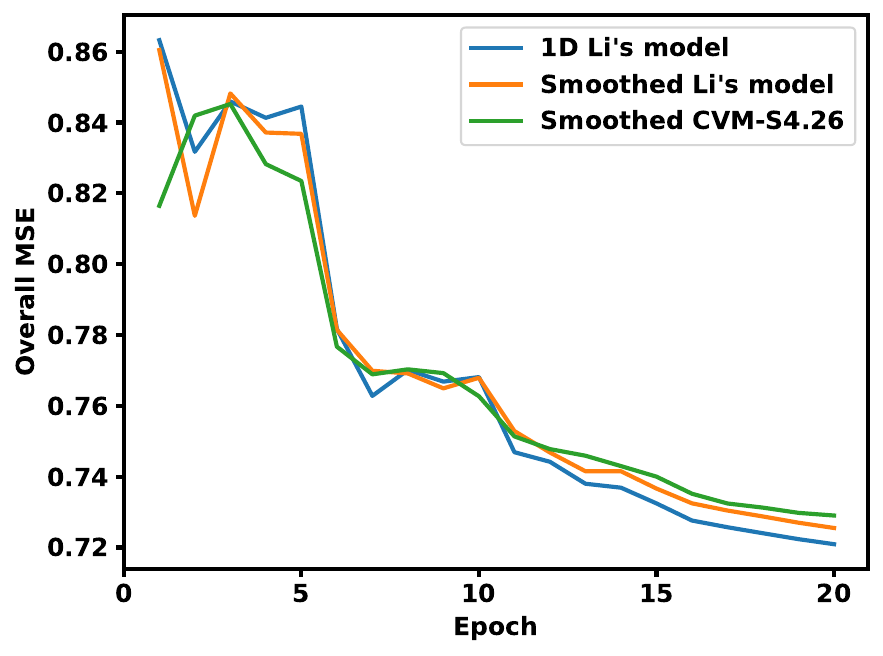}
\caption{Overall MSE (mean squared error) in the FWI process for SB1 with different initial models, corresponding to Figure \ref{real}. Higher frequency data is progressively incorporated, as in synthetic tests. We use an Adam optimizer with a learning rate of $0.05$ and a scheduler that decays the learning rate by half every $5$ epochs. The batch size is $64$.}
\label{ltoh_real}
\end{figure}

\begin{figure}
\centering
\includegraphics[width=1.0\textwidth]{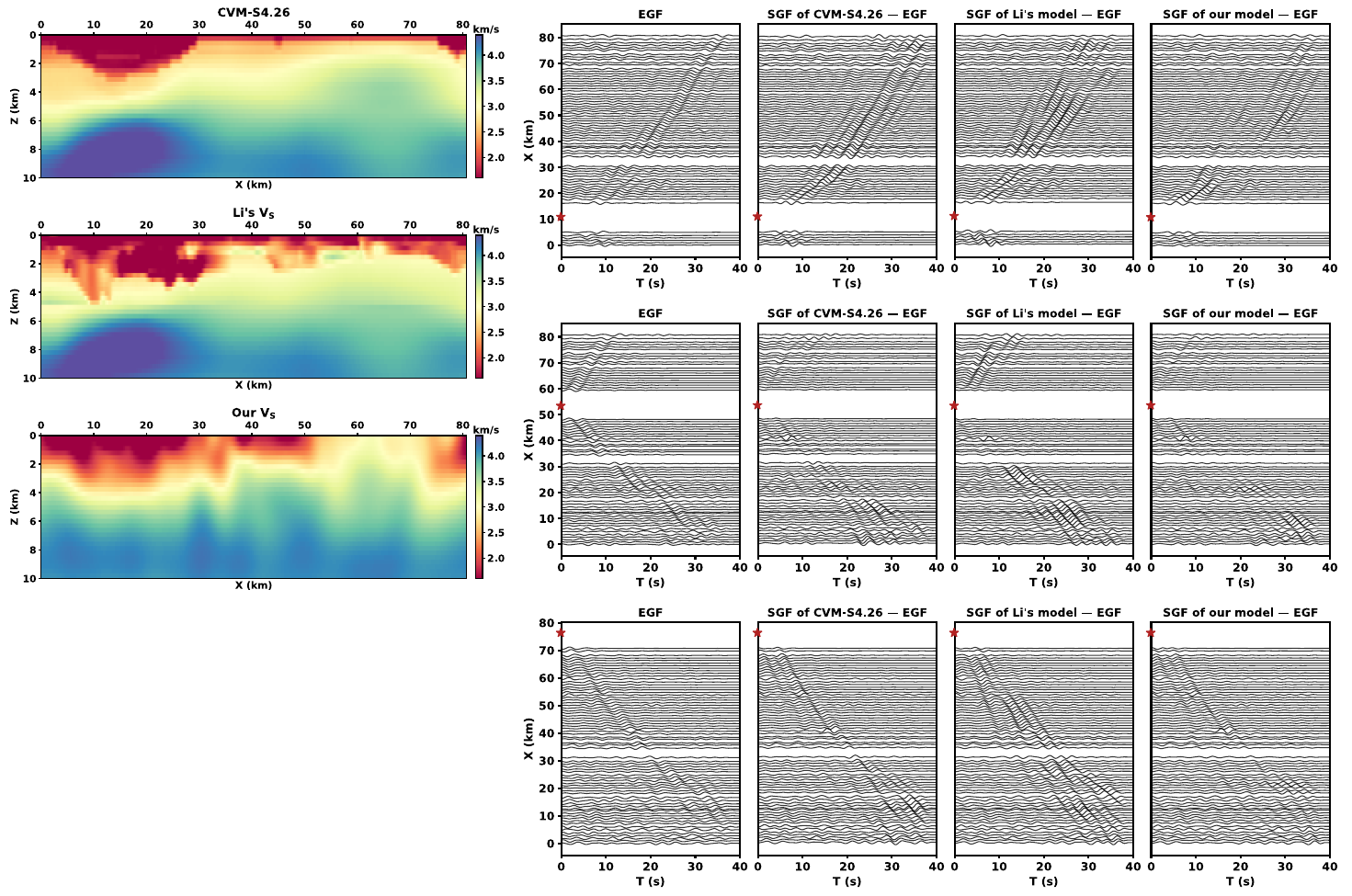}
\caption{Waveform difference plots corresponding to Figure \ref{waveform}.}
\label{waveform_diff}
\end{figure}

\newpage
\begin{table*}
\centering
\caption{Details in each neural operator layer.}
\label{hyperparam}
\begin{tabular}{@{}llllll}
\hline
Layer & Input co-dimension & Output co-dimension & Dimension & Number of modes & Activation\\
\hline
FNO 1 & 32 (lifting dimension) & 64 & (192, 96) & (96, 48) & GELU \\
FNO 2 & 64 & 128 & (128, 64) & (64, 32) & GELU \\
FNO 3 & 128 & 256 & (64, 32) & (32, 16) & GELU \\
FNO 4 & 256 & 512 & (32, 16) & (16, 8) & GELU \\
FNO 5 & 512 & 256 & (64, 32) & (16, 8) & GELU \\
FNO 6 & 512 & 128 & (128, 64) & (32, 16) & GELU \\
FNO 7 & 256 & 64 & (192, 96) & (64, 32) & GELU \\
FNO 8 & 128 & 64 & (256, 128) & (96, 48) & GELU \\
\hline
Layer & Input channel & Output channel & Dimension & Hidden neurons & Activation\\
\hline
GNO & 100 & 96 & (256, 1)& (96, 96) & GELU \\
\hline
\end{tabular}
\end{table*}

\begin{table*}
\centering
\caption{Parameters for the random fields used to generate velocity models.}
\label{rf}
\begin{tabular}{@{}llllll}
\hline
Velocity model & Covariance & $var$ & $len\_scale$ & $nu$ & Background\\
\hline
$V_S=BG_S\times (1+RF_S/100)$ & $RF_S$: Matérn & 400 & (64, 8) & 1.5 & $BG_S$: 1D from \cite{lee2014full} \\
$V_P=BG_P\times (1+RF_P/100)$ & $RF_P$: Matérn & 4 & (64, 8) & 2.5 & $BG_P(V_S)$: \cite{brocher2005empirical}\\
\hline
\end{tabular}
\end{table*}

%% file: main.bib
@article{maguire2022magma,
  title={{Magma accumulation at depths of prior rhyolite storage beneath Yellowstone Caldera}},
  author={Maguire, Ross and Schmandt, Brandon and Li, Jiaqi and Jiang, Chengxin and Li, Guoliang and Wilgus, Justin and Chen, Min},
  journal={Science},
  volume={378},
  number={6623},
  pages={1001--1004},
  year={2022},
  publisher={American Association for the Advancement of Science}
}

@article{liu20173,
  title={{3-D crustal and uppermost mantle structure beneath NE China revealed by ambient noise adjoint tomography}},
  author={Liu, Yaning and Niu, Fenglin and Chen, Min and Yang, Wencai},
  journal={Earth and Planetary Science Letters},
  volume={461},
  pages={20--29},
  year={2017},
  publisher={Elsevier}
}

@article{wapenaar2004retrieving,
  title={{Retrieving the Elastodynamic Green's Function of an Arbitrary Inhomogeneous Medium by Cross Correlation}},
  author={Wapenaar, Kees},
  journal={Physical review letters},
  volume={93},
  number={25},
  pages={254301},
  year={2004},
  publisher={APS}
}

@article{guo2015high,
  title={{High resolution 3-D crustal structure beneath NE China from joint inversion of ambient noise and receiver functions using NECESSArray data}},
  author={Guo, Zhen and Chen, Y John and Ning, Jieyuan and Feng, Yongge and Grand, Stephen P and Niu, Fenglin and Kawakatsu, Hitoshi and Tanaka, Satoru and Obayashi, Masayuki and Ni, James},
  journal={Earth and Planetary Science Letters},
  volume={416},
  pages={1--11},
  year={2015},
  publisher={Elsevier}
}

@article{lin2008surface,
  title={{Surface wave tomography of the western United States from ambient seismic noise: Rayleigh and Love wave phase velocity maps}},
  author={Lin, Fan-Chi and Moschetti, Morgan P and Ritzwoller, Michael H},
  journal={Geophysical Journal International},
  volume={173},
  number={1},
  pages={281--298},
  year={2008},
  publisher={Blackwell Publishing Ltd Oxford, UK}
}

@article{virieux2009overview,
  title={{An overview of full-waveform inversion in exploration geophysics}},
  author={Virieux, Jean and Operto, St{\'e}phane},
  journal={Geophysics},
  volume={74},
  number={6},
  pages={WCC1--WCC26},
  year={2009},
  publisher={Society of Exploration Geophysicists}
}

@article{plessix2006review,
  title={{A review of the adjoint-state method for computing the gradient of a functional with geophysical applications}},
  author={Plessix, R-E},
  journal={Geophysical Journal International},
  volume={167},
  number={2},
  pages={495--503},
  year={2006},
  publisher={Blackwell Publishing Ltd Oxford, UK}
}

@article{zou2024deep,
  title={{Deep neural Helmholtz operators for 3-D elastic wave propagation and inversion}},
  author={Zou, Caifeng and Azizzadenesheli, Kamyar and Ross, Zachary E and Clayton, Robert W},
  journal={Geophysical Journal International},
  volume={239},
  number={3},
  pages={1469--1484},
  year={2024},
  publisher={Oxford University Press}
}

@article{yang2023rapid,
  title={{Rapid seismic waveform modeling and inversion with neural operators}},
  author={Yang, Yan and Gao, Angela F and Azizzadenesheli, Kamyar and Clayton, Robert W and Ross, Zachary E},
  journal={IEEE Transactions on Geoscience and Remote Sensing},
  volume={61},
  pages={1--12},
  year={2023},
  publisher={IEEE}
}

@article{li2020neural,
  title={{Neural operator: Graph kernel network for partial differential equations}},
  author={Li, Zongyi and Kovachki, Nikola and Azizzadenesheli, Kamyar and Liu, Burigede and Bhattacharya, Kaushik and Stuart, Andrew and Anandkumar, Anima},
  journal={arXiv preprint arXiv:2003.03485},
  year={2020}
}

@article{li2020fourier,
  title={{Fourier neural operator for parametric partial differential equations}},
  author={Li, Zongyi and Kovachki, Nikola and Azizzadenesheli, Kamyar and Liu, Burigede and Bhattacharya, Kaushik and Stuart, Andrew and Anandkumar, Anima},
  journal={arXiv preprint arXiv:2010.08895},
  year={2020}
}

@article{raissi2019physics,
  title={{Physics-informed neural networks: A deep learning framework for solving forward and inverse problems involving nonlinear partial differential equations}},
  author={Raissi, Maziar and Perdikaris, Paris and Karniadakis, George E},
  journal={Journal of Computational physics},
  volume={378},
  pages={686--707},
  year={2019},
  publisher={Elsevier}
}

@article{baydin2018automatic,
  title={{Automatic differentiation in machine learning: a survey}},
  author={Baydin, Atilim Gunes and Pearlmutter, Barak A and Radul, Alexey Andreyevich and Siskind, Jeffrey Mark},
  journal={Journal of machine learning research},
  volume={18},
  number={153},
  pages={1--43},
  year={2018}
}

@article{wu2019inversionnet,
  title={{InversionNet: An efficient and accurate data-driven full waveform inversion}},
  author={Wu, Yue and Lin, Youzuo},
  journal={IEEE Transactions on Computational Imaging},
  volume={6},
  pages={419--433},
  year={2019},
  publisher={IEEE}
}

@article{moseley2020deep,
  title={{Deep learning for fast simulation of seismic waves in complex media}},
  author={Moseley, Ben and Nissen-Meyer, Tarje and Markham, Andrew},
  journal={Solid Earth},
  volume={11},
  number={4},
  pages={1527--1549},
  year={2020},
  publisher={Copernicus Publications G{\"o}ttingen, Germany}
}

@article{panning2009seismic,
  title={{Seismic waveform modelling in a 3-D Earth using the Born approximation: potential shortcomings and a remedy}},
  author={Panning, Mark P and Capdeville, Yann and Romanowicz, Barbara A},
  journal={Geophysical Journal International},
  volume={177},
  number={1},
  pages={161--178},
  year={2009},
  publisher={Blackwell Publishing Ltd Oxford, UK}
}

@article{zhang2018linear,
  title={{Linear array ambient noise adjoint tomography reveals intense crust-mantle interactions in North China Craton}},
  author={Zhang, Chao and Yao, Huajian and Liu, Qinya and Zhang, Ping and Yuan, Yanhua O and Feng, Jikun and Fang, Lihua},
  journal={Journal of Geophysical Research: Solid Earth},
  volume={123},
  number={1},
  pages={368--383},
  year={2018},
  publisher={Wiley Online Library}
}

@article{devito,
  author = {Louboutin, M. and Lange, M. and Luporini, F. and Kukreja, N. and Witte, P. A. and Herrmann, F. J. and Velesko, P. and Gorman, G. J.},
  title = {{Devito (v3.1.0): an embedded domain-specific language for finite differences and geophysical exploration}},
  journal = {Geoscientific Model Development},
  volume = {12},
  year = {2019},
  number = {3},
  pages = {1165--1187},
  url = {https://www.geosci-model-dev.net/12/1165/2019/},
  doi = {10.5194/gmd-12-1165-2019}
}

@article{tromp2005seismic,
  title={{Seismic tomography, adjoint methods, time reversal and banana-doughnut kernels}},
  author={Tromp, Jeroen and Tape, Carl and Liu, Qinya},
  journal={Geophysical Journal International},
  volume={160},
  number={1},
  pages={195--216},
  year={2005},
  publisher={Blackwell Publishing Ltd Oxford, UK}
}

@article{hornik1989multilayer,
  title={{Multilayer feedforward networks are universal approximators}},
  author={Hornik, Kurt and Stinchcombe, Maxwell and White, Halbert},
  journal={Neural networks},
  volume={2},
  number={5},
  pages={359--366},
  year={1989},
  publisher={Elsevier}
}

@article{Kovachki,
  title={{Neural Operator: Learning Maps Between Function Spaces With Applications to PDEs.}},
  author={Kovachki, Nikola B and Li, Zongyi and Liu, Burigede and Azizzadenesheli, Kamyar and Bhattacharya, Kaushik and Stuart, Andrew M and Anandkumar, Anima},
  journal={J. Mach. Learn. Res.},
  volume={24},
  number={89},
  pages={1--97},
  year={2023}
}

@article{Rahman,
  title={{U-NO: U-shaped neural operators}},
  author={Rahman, Md Ashiqur and Ross, Zachary E and Azizzadenesheli, Kamyar},
  journal={arXiv preprint arXiv:2204.11127},
  year={2022}
}

@inproceedings{Ronneberger,
  title={{U-net: Convolutional networks for biomedical image segmentation}},
  author={Ronneberger, Olaf and Fischer, Philipp and Brox, Thomas},
  booktitle={Medical Image Computing and Computer-Assisted Intervention--MICCAI 2015: 18th International Conference, Munich, Germany, October 5-9, 2015, Proceedings, Part III 18},
  pages={234--241},
  year={2015},
  organization={Springer}
}

@article{rumelhart1986learning,
  title={{Learning representations by back-propagating errors}},
  author={Rumelhart, David E and Hinton, Geoffrey E and Williams, Ronald J},
  journal={nature},
  volume={323},
  number={6088},
  pages={533--536},
  year={1986},
  publisher={Nature Publishing Group UK London}
}

@article{elliott2018simple,
  title={{The simple essence of automatic differentiation}},
  author={Elliott, Conal},
  journal={Proceedings of the ACM on Programming Languages},
  volume={2},
  number={ICFP},
  pages={1--29},
  year={2018},
  publisher={ACM New York, NY, USA}
}

@article{Afanasiev,
  title={{Modular and flexible spectral-element waveform modelling in two and three dimensions}},
  author={Afanasiev, Michael and Boehm, Christian and van Driel, Martin and Krischer, Lion and Rietmann, Max and May, Dave A and Knepley, Matthew G and Fichtner, Andreas},
  journal={Geophysical Journal International},
  volume={216},
  number={3},
  pages={1675--1692},
  year={2019},
  publisher={Oxford University Press}
}

@article{brocher2005empirical,
  title={{Empirical relations between elastic wavespeeds and density in the Earth's crust}},
  author={Brocher, Thomas M},
  journal={Bulletin of the seismological Society of America},
  volume={95},
  number={6},
  pages={2081--2092},
  year={2005},
  publisher={Seismological Society of America}
}

@article{lee2014full,
  title={{Full-3-D tomography for crustal structure in southern California based on the scattering-integral and the adjoint-wavefield methods}},
  author={Lee, En-Jui and Chen, Po and Jordan, Thomas H and Maechling, Phillip B and Denolle, Marine AM and Beroza, Gregory C},
  journal={Journal of Geophysical Research: Solid Earth},
  volume={119},
  number={8},
  pages={6421--6451},
  year={2014},
  publisher={Wiley Online Library}
}

@article{clayton2019exposing,
  title={{Exposing Los Angeles’s Shaky Geologic Underbelly (Vol. 100)}},
  author={Clayton, R and Persaud, P and Denolle, M and Polet, J},
  journal={Eos. https://doi. org/10.1029},
  year={2019}
}

@article{Kingma,
  title={{Adam: A method for stochastic optimization}},
  author={Kingma, Diederik P and Ba, Jimmy},
  journal={arXiv preprint arXiv:1412.6980},
  year={2014}
}

@incollection{wright1991structural,
  author       = {Wright, T. L.},
  title        = {{Structural geology and tectonic evolution of the Los Angeles basin, California}},
  booktitle    = {Active Margin Basins},
  editor       = {Biddle, K. T.},
  volume       = {52},
  series       = {AAPG Memoir},
  pages        = {35--79},
  year         = {1991},
  publisher    = {American Association of Petroleum Geologists},
}

@article{villa2023three,
  title={{Three-Dimensional Basin Depth Map of the Northern Los Angeles Basins From Gravity and Seismic Measurements}},
  author={Villa, Valeria and Li, Yida and Clayton, Robert W and Persaud, Patricia},
  journal={Journal of Geophysical Research: Solid Earth},
  volume={128},
  number={7},
  pages={e2022JB025425},
  year={2023},
  publisher={Wiley Online Library}
}

@article{ghose2023basin,
  title={{Basin Structure for Earthquake Ground Motion Estimates in Urban Los Angeles Mapped with Nodal Receiver Functions}},
  author={Ghose, Ritu and Persaud, Patricia and Clayton, Robert W},
  journal={Geosciences},
  volume={13},
  number={11},
  pages={320},
  year={2023},
  publisher={MDPI}
}

@article{chen2014low,
  title={{Low wave speed zones in the crust beneath SE Tibet revealed by ambient noise adjoint tomography}},
  author={Chen, Min and Huang, Hui and Yao, Huajian and van der Hilst, Rob and Niu, Fenglin},
  journal={Geophysical Research Letters},
  volume={41},
  number={2},
  pages={334--340},
  year={2014},
  publisher={Wiley Online Library}
}

@article{liu2006finite,
  title={{Finite-frequency kernels based on adjoint methods}},
  author={Liu, Qinya and Tromp, Jeroen},
  journal={Bulletin of the Seismological Society of America},
  volume={96},
  number={6},
  pages={2383--2397},
  year={2006},
  publisher={Seismological Society of America}
}

@article{wang2021adjoint,
  title={{Adjoint tomography of ambient noise data and teleseismic P waves: Methodology and applications to central California}},
  author={Wang, Kai and Yang, Yingjie and Jiang, Chengxin and Wang, Yi and Tong, Ping and Liu, Tianshi and Liu, Qinya},
  journal={Journal of Geophysical Research: Solid Earth},
  volume={126},
  number={6},
  pages={e2021JB021648},
  year={2021},
  publisher={Wiley Online Library}
}

@article{wang2018refined,
  title={{Refined crustal and uppermost mantle structure of southern California by ambient noise adjoint tomography}},
  author={Wang, Kai and Yang, Yingjie and Basini, Piero and Tong, Ping and Tape, Carl and Liu, Qinya},
  journal={Geophysical Journal International},
  volume={215},
  number={2},
  pages={844--863},
  year={2018},
  publisher={Oxford University Press}
}

@article{yao2010heterogeneity,
  title={{Heterogeneity and anisotropy of the lithosphere of SE Tibet from surface wave array tomography}},
  author={Yao, Huajian and van Der Hilst, Robert D and Montagner, Jean-Paul},
  journal={Journal of Geophysical Research: Solid Earth},
  volume={115},
  number={B12},
  year={2010},
  publisher={Wiley Online Library}
}

@article{bensen2007processing,
  title={{Processing seismic ambient noise data to obtain reliable broad-band surface wave dispersion measurements}},
  author={Bensen, GD and Ritzwoller, MH and Barmin, MP and Levshin, A Lin and Lin, Feifan and Moschetti, MP and Shapiro, NM and Yang, Yanyan},
  journal={Geophysical journal international},
  volume={169},
  number={3},
  pages={1239--1260},
  year={2007},
  publisher={Blackwell Publishing Ltd Oxford, UK}
}

@article{li2023shear,
  title={{Shear Wave Velocities in the San Gabriel and San Bernardino Basins, California}},
  author={Li, Yida and Villa, Valeria and Clayton, Robert W and Persaud, Patricia},
  journal={Journal of Geophysical Research: Solid Earth},
  volume={128},
  number={7},
  pages={e2023JB026488},
  year={2023},
  publisher={Wiley Online Library}
}

@article{forbriger2014line,
  title={{Line-source simulation for shallow-seismic data. Part 1: theoretical background}},
  author={Forbriger, Thomas and Groos, Lisa and Sch{\"a}fer, Martin},
  journal={Geophysical Journal International},
  volume={198},
  number={3},
  pages={1387--1404},
  year={2014},
  publisher={Oxford University Press}
}

@article{schafer2014line,
  title={{Line-source simulation for shallow-seismic data. Part 2: full-waveform inversion—a synthetic 2-D case study}},
  author={Sch{\"a}fer, Martin and Groos, Lisa and Forbriger, Thomas and Bohlen, Thomas},
  journal={Geophysical Journal International},
  volume={198},
  number={3},
  pages={1405--1418},
  year={2014},
  publisher={Oxford University Press}
}

@article{yosinski2014transferable,
  title={{How transferable are features in deep neural networks?}},
  author={Yosinski, Jason and Clune, Jeff and Bengio, Yoshua and Lipson, Hod},
  journal={Advances in neural information processing systems},
  volume={27},
  year={2014}
}

@article{Zou2024,
    author = {Zou, Caifeng and Clayton, Robert W.},
    title = {{Imaging the Northern Los Angeles Basins with Autocorrelations}},
    journal = {Seismological Research Letters},
    volume = {96},
    number = {3},
    pages = {1791-1801},
    year = {2024},
    month = {11},
    issn = {0895-0695},
    doi = {10.1785/0220240140},
    url = {https://doi.org/10.1785/0220240140},
    eprint = {https://pubs.geoscienceworld.org/ssa/srl/article-pdf/96/3/1791/7038288/srl-2024140.1.pdf},
}

@article{tape2010seismic,
  title={{Seismic tomography of the southern California crust based on spectral-element and adjoint methods}},
  author={Tape, Carl and Liu, Qinya and Maggi, Alessia and Tromp, Jeroen},
  journal={Geophysical Journal International},
  volume={180},
  number={1},
  pages={433--462},
  year={2010},
  publisher={Blackwell Publishing Ltd Oxford, UK}
}

@article{zhu2015seismic,
  title={{Seismic structure of the European upper mantle based on adjoint tomography}},
  author={Zhu, Hejun and Bozda{\u{g}}, Ebru and Tromp, Jeroen},
  journal={Geophysical Journal International},
  volume={201},
  number={1},
  pages={18--52},
  year={2015},
  publisher={Oxford University Press}
}

@article{fournier2019inversion,
  title={{Inversion using spatially variable mixed Lp norms}},
  author={Fournier, Dominique and Oldenburg, Douglas W},
  journal={Geophysical Journal International},
  volume={218},
  number={1},
  pages={268--282},
  year={2019},
  publisher={Oxford University Press}
}

@article{sambridge2022geophysical,
  title={{Geophysical inversion and optimal transport}},
  author={Sambridge, Malcolm and Jackson, Andrew and Valentine, Andrew P},
  journal={Geophysical Journal International},
  volume={231},
  number={1},
  pages={172--198},
  year={2022},
  publisher={Oxford University Press}
}

@article{li2024physics,
  title={{Physics-informed neural operator for learning partial differential equations}},
  author={Li, Zongyi and Zheng, Hongkai and Kovachki, Nikola and Jin, David and Chen, Haoxuan and Liu, Burigede and Azizzadenesheli, Kamyar and Anandkumar, Anima},
  journal={ACM/JMS Journal of Data Science},
  volume={1},
  number={3},
  pages={1--27},
  year={2024},
  publisher={ACM New York, NY}
}

@article{Rasht‐Behesht,
  title={{Physics-informed neural networks (PINNs) for wave propagation and full waveform inversions}},
  author={Rasht-Behesht, Majid and Huber, Christian and Shukla, Khemraj and Karniadakis, George Em},
  journal={Journal of Geophysical Research: Solid Earth},
  volume={127},
  number={5},
  pages={e2021JB023120},
  year={2022},
  publisher={Wiley Online Library}
}

@article{ren2024seismicnet,
  title={{SeismicNet: Physics-informed neural networks for seismic wave modeling in semi-infinite domain}},
  author={Ren, Pu and Rao, Chengping and Chen, Su and Wang, Jian-Xun and Sun, Hao and Liu, Yang},
  journal={Computer Physics Communications},
  volume={295},
  pages={109010},
  year={2024},
  publisher={Elsevier}
}

@article{SongAlkhalifah,
  title={{Wavefield reconstruction inversion via physics-informed neural networks}},
  author={Song, Chao and Alkhalifah, Tariq A},
  journal={IEEE Transactions on Geoscience and Remote Sensing},
  volume={60},
  pages={1--12},
  year={2021},
  publisher={IEEE}
}

@article{bozdaug2011misfit,
  title={{Misfit functions for full waveform inversion based on instantaneous phase and envelope measurements}},
  author={Bozda{\u{g}}, Ebru and Trampert, Jeannot and Tromp, Jeroen},
  journal={Geophysical Journal International},
  volume={185},
  number={2},
  pages={845--870},
  year={2011},
  publisher={Blackwell Publishing Ltd Oxford, UK}
}

@article{azizzadenesheli2024neural,
  title={{Neural operators for accelerating scientific simulations and design}},
  author={Azizzadenesheli, Kamyar and Kovachki, Nikola and Li, Zongyi and Liu-Schiaffini, Miguel and Kossaifi, Jean and Anandkumar, Anima},
  journal={Nature Reviews Physics},
  volume={6},
  number={5},
  pages={320--328},
  year={2024},
  publisher={Nature Publishing Group UK London}
}

@article{yang2021seismic,
  title={{Seismic wave propagation and inversion with neural operators}},
  author={Yang, Yan and Gao, Angela F and Castellanos, Jorge C and Ross, Zachary E and Azizzadenesheli, Kamyar and Clayton, Robert W},
  journal={The Seismic Record},
  volume={1},
  number={3},
  pages={126--134},
  year={2021},
  publisher={Seismological Society of America}
}

@article{li2020multipole,
  title={{Multipole graph neural operator for parametric partial differential equations}},
  author={Li, Zongyi and Kovachki, Nikola and Azizzadenesheli, Kamyar and Liu, Burigede and Stuart, Andrew and Bhattacharya, Kaushik and Anandkumar, Anima},
  journal={Advances in Neural Information Processing Systems},
  volume={33},
  pages={6755--6766},
  year={2020}
}

@article{hendrycks2016gaussian,
  title={{Gaussian error linear units (gelus)}},
  author={Hendrycks, Dan and Gimpel, Kevin},
  journal={arXiv preprint arXiv:1606.08415},
  year={2016}
}

@article{sager2020global,
  title={{Global-scale full-waveform ambient noise inversion}},
  author={Sager, Korbinian and Boehm, Christian and Ermert, Laura and Krischer, Lion and Fichtner, Andreas},
  journal={Journal of Geophysical Research: Solid Earth},
  volume={125},
  number={4},
  pages={e2019JB018644},
  year={2020},
  publisher={Wiley Online Library}
}

@article{sager2018towards,
  title={{Towards full waveform ambient noise inversion}},
  author={Sager, Korbinian and Ermert, Laura and Boehm, Christian and Fichtner, Andreas},
  journal={Geophysical Journal International},
  volume={212},
  number={1},
  pages={566--590},
  year={2018},
  publisher={Oxford University Press}
}

@article{tsai2024towards,
  title={{Towards limited-domain full waveform ambient noise inversion}},
  author={Tsai, Victor C and Sager, Korbinian and Bowden, Daniel C},
  journal={Geophysical Journal International},
  volume={237},
  number={2},
  pages={965--973},
  year={2024},
  publisher={Oxford University Press}
}

@article{tromp2010noise,
  title={{Noise cross-correlation sensitivity kernels}},
  author={Tromp, Jeroen and Luo, Yang and Hanasoge, Shravan and Peter, Daniel},
  journal={Geophysical Journal International},
  volume={183},
  number={2},
  pages={791--819},
  year={2010},
  publisher={Blackwell Publishing Ltd Oxford, UK}
}

@article{liu2023ambient,
  title={{Ambient noise differential adjoint tomography reveals fluid-bearing rocks near active faults in Los Angeles}},
  author={Liu, Xin and Beroza, Gregory C and Li, Hongyi},
  journal={Nature Communications},
  volume={14},
  number={1},
  pages={6873},
  year={2023},
  publisher={Nature Publishing Group UK London}
}

@article{zheng2011crust,
  title={{Crust and uppermost mantle beneath the North China Craton, northeastern China, and the Sea of Japan from ambient noise tomography}},
  author={Zheng, Yong and Shen, Weisen and Zhou, Longquan and Yang, Yingjie and Xie, Zujun and Ritzwoller, Michael H},
  journal={Journal of Geophysical Research: Solid Earth},
  volume={116},
  number={B12},
  year={2011},
  publisher={Wiley Online Library}
}

@article{fichtner2006adjoint1,
  title={{The adjoint method in seismology: I. Theory}},
  author={Fichtner, Andreas and Bunge, H-P and Igel, Heiner},
  journal={Physics of the Earth and Planetary Interiors},
  volume={157},
  number={1-2},
  pages={86--104},
  year={2006},
  publisher={Elsevier}
}

@article{fichtner2006adjoint2,
  title={{The adjoint method in seismology—: II. Applications: Traveltimes and sensitivity functionals}},
  author={Fichtner, A and Bunge, H-P and Igel, H},
  journal={Physics of the Earth and Planetary Interiors},
  volume={157},
  number={1-2},
  pages={105--123},
  year={2006},
  publisher={Elsevier}
}

@article{wapenaar2010tutorial,
  title={{Tutorial on seismic interferometry: Part 2—Underlying theory and new advances}},
  author={Wapenaar, Kees and Slob, Evert and Snieder, Roel and Curtis, Andrew},
  journal={Geophysics},
  volume={75},
  number={5},
  pages={75A211--75A227},
  year={2010},
  publisher={Society of Exploration Geophysicists}
}

@article{beydoun1988first,
  title={{First Born and Rytov approximations: Modeling and inversion conditions in a canonical example}},
  author={Beydoun, Wafik B and Tarantola, Albert},
  journal={The Journal of the Acoustical Society of America},
  volume={83},
  number={3},
  pages={1045--1055},
  year={1988},
  publisher={Acoustical Society of America}
}

@article{pladys2021cycle,
  title={{On cycle-skipping and misfit function modification for full-wave inversion: Comparison of five recent approaches}},
  author={Pladys, Arnaud and Brossier, Romain and Li, Yubing and M{\'e}tivier, Ludovic},
  journal={Geophysics},
  volume={86},
  number={4},
  pages={R563--R587},
  year={2021},
  publisher={Society of Exploration Geophysicists}
}

@article{gauthier1986two,
  title={{Two-dimensional nonlinear inversion of seismic waveforms: Numerical results}},
  author={Gauthier, Odile and Virieux, Jean and Tarantola, Albert},
  journal={geophysics},
  volume={51},
  number={7},
  pages={1387--1403},
  year={1986},
  publisher={Society of Exploration Geophysicists}
}

@article{metivier2018optimal,
  title={{Optimal transport for mitigating cycle skipping in full-waveform inversion: A graph-space transform approach}},
  author={M{\'e}tivier, Ludovic and Allain, Aude and Brossier, Romain and M{\'e}rigot, Quentin and Oudet, Edouard and Virieux, Jean},
  journal={Geophysics},
  volume={83},
  number={5},
  pages={R515--R540},
  year={2018},
  publisher={Society of Exploration Geophysicists}
}

@misc{caifeng_zou_2025_15061657,
  author       = {Caifeng Zou},
  title        = {{caifeng-zou/ANFWI\_HNO: First release}},
  month        = mar,
  year         = 2025,
  publisher    = {Zenodo},
  version      = {v1.0.0},
  doi          = {10.5281/zenodo.15061657},
  url          = {https://doi.org/10.5281/zenodo.15061657},
  howpublished = {Software},
  swhid        = {swh:1:dir:e9ceffdddcb9a17aa24807eca889ccadb9786f26
                   ;origin=https://doi.org/10.5281/zenodo.15061656;vi
                   sit=swh:1:snp:c8d79cf94e1b11290db56ecf29df79b4c520
                   baa4;anchor=swh:1:rel:f7f58120a607fe884f292c812747
                   1a19988fd0f2;path=caifeng-zou-ANFWI\_HNO-ba21e45
                  },
}

@misc{BASIN2018,
  author = "{BASIN}",
  year = {2018},
  title = {{San Gabriel and San Bernardino Basin Arrays}},
  organization = {Caltech},
  howpublished = {Dataset},
  doi = {10.7909/46a0-ma59}
}

@article{kong2025reducing,
  title={{Reducing Frequency Bias of Fourier Neural Operators in 3D Seismic Wavefield Simulations Through Multi-Stage Training}},
  author={Kong, Qingkai and Zou, Caifeng and Choi, Youngsoo and Matzel, Eric M and Azizzadenesheli, Kamyar and Ross, Zachary E and Rodgers, Arthur J and Clayton, Robert W},
  journal={arXiv preprint arXiv:2503.02023},
  year={2025}
}

@article{zhang2021rayleigh,
  title={{Rayleigh wave dispersion spectrum inversion across scales}},
  author={Zhang, Zhen-dong and Saygin, Erdinc and He, Leiyu and Alkhalifah, Tariq},
  journal={Surveys in Geophysics},
  volume={42},
  pages={1281--1303},
  year={2021},
  publisher={Springer},
  doi={10.1007/s10712-021-09667-z}
}

@article{zhang2020wave,
  title={{Wave-equation dispersion spectrum inversion for near-surface characterization using fibre-optics acquisition}},
  author={Zhang, Zhen-dong and Alajami, Mamdoh and Alkhalifah, Tariq},
  journal={Geophysical Journal International},
  volume={222},
  number={2},
  pages={907--918},
  year={2020},
  publisher={Oxford University Press}
}

@article{sun2019application,
  title={{The application of an optimal transport to a preconditioned data matching function for robust waveform inversion}},
  author={Sun, Bingbing and Alkhalifah, Tariq},
  journal={Geophysics},
  volume={84},
  number={6},
  pages={R923--R945},
  year={2019},
  publisher={Society of Exploration Geophysicists}
}

@book{Tariq2016,
    author = {Alkhalifah, Tariq A.},
    title = {{Full waveform inversion in an anisotropic world: Where are the parameters hiding?}},
    publisher = {EAGE},
    year = {2016},
    month = {01},
    isbn = {9789462821927},
    doi = {10.3997/9789462822023},
    url = {https://doi.org/10.3997/9789462822023},
}

@article{li2023solving,
  title={{Solving seismic wave equations on variable velocity models with Fourier neural operator}},
  author={Li, Bian and Wang, Hanchen and Feng, Shihang and Yang, Xiu and Lin, Youzuo},
  journal={IEEE Transactions on Geoscience and Remote Sensing},
  volume={61},
  pages={1--18},
  year={2023},
  publisher={IEEE}
}

@article{zhang2023learning,
  title={{Learning to solve the elastic wave equation with Fourier neural operators}},
  author={Zhang, Tianze and Trad, Daniel and Innanen, Kristopher},
  journal={Geophysics},
  volume={88},
  number={3},
  pages={T101--T119},
  year={2023},
  publisher={Society of Exploration Geophysicists}
}

@article{wei2022small,
  title={{Small-data-driven fast seismic simulations for complex media using physics-informed Fourier neural operators}},
  author={Wei, Wei and Fu, Li-Yun},
  journal={Geophysics},
  volume={87},
  number={6},
  pages={T435--T446},
  year={2022},
  publisher={Society of Exploration Geophysicists}
}

@article{lehmann20243d,
  title={{3D elastic wave propagation with a factorized Fourier neural operator (F-FNO)}},
  author={Lehmann, Fanny and Gatti, Filippo and Bertin, Micha{\"e}l and Clouteau, Didier},
  journal={Computer Methods in Applied Mechanics and Engineering},
  volume={420},
  pages={116718},
  year={2024},
  publisher={Elsevier}
}

@article{lehmann2025multiple,
  title={{Multiple-input fourier neural operator (mifno) for source-dependent 3d elastodynamics}},
  author={Lehmann, Fanny and Gatti, Filippo and Clouteau, Didier},
  journal={Journal of Computational Physics},
  pages={113813},
  year={2025},
  publisher={Elsevier}
}

@article{huang2025learned,
  title={{Learned frequency-domain scattered wavefield solutions using neural operators}},
  author={Huang, Xinquan and Alkhalifah, Tariq},
  journal={Geophysical Journal International},
  volume={241},
  number={3},
  pages={1467--1478},
  year={2025},
  publisher={Oxford University Press}
}

@article{li2025feature,
  title={{A Feature Enhanced Autoencoder Integrated with Fourier Neural Operator for Intelligent Elastic Wavefield Modeling}},
  author={Li, Chen and Zhao, Haixia and Hao, Yufan},
  journal={IEEE Transactions on Geoscience and Remote Sensing},
  year={2025},
  publisher={IEEE}
}

@article{cheng2025seismic,
  title={{Seismic wavefield solutions via physics-guided generative neural operator}},
  author={Cheng, Shijun and Taufik, Mohammad H and Alkhalifah, Tariq},
  journal={arXiv preprint arXiv:2503.06488},
  year={2025}
}

@inproceedings{LeCun,
  title={{A theoretical framework for back-propagation}},
  author={LeCun, Yann and Touresky, D and Hinton, G and Sejnowski, T},
  booktitle={Proceedings of the 1988 connectionist models summer school},
  volume={1},
  pages={21--28},
  year={1988},
  organization={San Mateo, CA, USA}
}

@article{Zhu,
  title={{A general approach to seismic inversion with automatic differentiation}},
  author={Zhu, Weiqiang and Xu, Kailai and Darve, Eric and Beroza, Gregory C},
  journal={Computers \& Geosciences},
  volume={151},
  pages={104751},
  year={2021},
  publisher={Elsevier}
}

@inbook{pytorch,
author = {Paszke, Adam and Gross, Sam and Massa, Francisco and Lerer, Adam and Bradbury, James and Chanan, Gregory and Killeen, Trevor and Lin, Zeming and Gimelshein, Natalia and Antiga, Luca and Desmaison, Alban and K\"{o}pf, Andreas and Yang, Edward and DeVito, Zach and Raison, Martin and Tejani, Alykhan and Chilamkurthy, Sasank and Steiner, Benoit and Fang, Lu and Bai, Junjie and Chintala, Soumith},
title = {{PyTorch: an imperative style, high-performance deep learning library}},
year = {2019},
publisher = {Curran Associates Inc.},
address = {Red Hook, NY, USA},
booktitle = {Proceedings of the 33rd International Conference on Neural Information Processing Systems},
articleno = {721},
numpages = {12}
}

@incollection{lecun2002efficient,
  title={{Efficient backprop}},
  author={LeCun, Yann and Bottou, L{\'e}on and Orr, Genevieve B and M{\"u}ller, Klaus-Robert},
  booktitle={Neural networks: Tricks of the trade},
  pages={9--50},
  year={2002},
  publisher={Springer}
}

@article{van2020accelerated,
  title={{Accelerated full-waveform inversion using dynamic mini-batches}},
  author={van Herwaarden, Dirk Philip and Boehm, Christian and Afanasiev, Michael and Thrastarson, Solvi and Krischer, Lion and Trampert, Jeannot and Fichtner, Andreas},
  journal={Geophysical Journal International},
  volume={221},
  number={2},
  pages={1427--1438},
  year={2020},
  publisher={Oxford University Press}
}

@article{thrastarson2022data,
  title={{Data-adaptive global full-waveform inversion}},
  author={Thrastarson, Solvi and Van Herwaarden, Dirk-Philip and Krischer, Lion and Boehm, Christian and van Driel, Martin and Afanasiev, Michael and Fichtner, Andreas},
  journal={Geophysical Journal International},
  volume={230},
  number={2},
  pages={1374--1393},
  year={2022},
  publisher={Oxford University Press}
}

@article{bernal2021accelerating,
  title={{Accelerating full-waveform inversion through adaptive gradient optimization methods and dynamic simultaneous sources}},
  author={Bernal-Romero, Marcos and Iturrar{\'a}n-Viveros, Ursula},
  journal={Geophysical Journal International},
  volume={225},
  number={1},
  pages={97--126},
  year={2021},
  publisher={Oxford University Press}
}

@article{richardson2018seismic,
  title={{Seismic full-waveform inversion using deep learning tools and techniques}},
  author={Richardson, Alan},
  journal={arXiv preprint arXiv:1801.07232},
  year={2018}
}

@inproceedings{bottou2010large,
  title={{Large-scale machine learning with stochastic gradient descent}},
  author={Bottou, L{\'e}on},
  booktitle={Proceedings of COMPSTAT'2010: 19th International Conference on Computational StatisticsParis France, August 22-27, 2010 Keynote, Invited and Contributed Papers},
  pages={177--186},
  year={2010},
  organization={Springer}
}

@article{sun2020theory,
  title={{A theory-guided deep-learning formulation and optimization of seismic waveform inversion}},
  author={Sun, Jian and Niu, Zhan and Innanen, Kristopher A and Li, Junxiao and Trad, Daniel O},
  journal={Geophysics},
  volume={85},
  number={2},
  pages={R87--R99},
  year={2020},
  publisher={Society of Exploration Geophysicists}
}

@article{kazei2021mapping,
  title={{Mapping full seismic waveforms to vertical velocity profiles by deep learning}},
  author={Kazei, Vladimir and Ovcharenko, Oleg and Plotnitskii, Pavel and Peter, Daniel and Zhang, Xiangliang and Alkhalifah, Tariq},
  journal={Geophysics},
  volume={86},
  number={5},
  pages={R711--R721},
  year={2021},
  publisher={Society of Exploration Geophysicists}
}

@article{liu2020finite,
  title={{Finite-frequency sensitivity kernels for seismic noise interferometry based on differential time measurements}},
  author={Liu, Xin},
  journal={Journal of Geophysical Research: Solid Earth},
  volume={125},
  number={4},
  pages={e2019JB018932},
  year={2020},
  publisher={Wiley Online Library}
}

@ARTICLE{mao2025automatic,
  author={Mao, Shibo and Song, Peng and Tong, Siyou and Tan, Jun and Xie, Chuang and Zu, Guochang and Liu, Guangzhao},
  journal={IEEE Transactions on Geoscience and Remote Sensing}, 
  title={{Automatic Differentiation-Based Full Waveform Inversion of Anisotropic Parameters in TTI Media}}, 
  year={2025},
  volume={},
  number={},
  pages={1-1},
  doi={10.1109/TGRS.2025.3586294}}

@article{wang2021elastic,
  title={{Elastic isotropic and anisotropic full-waveform inversions using automatic differentiation for gradient calculations in a framework of recurrent neural networks}},
  author={Wang, Wenlong and McMechan, George A and Ma, Jianwei},
  journal={Geophysics},
  volume={86},
  number={6},
  pages={R795--R810},
  year={2021},
  publisher={Society of Exploration Geophysicists}
}

@article{liu2023robust,
  title={{Robust full-waveform inversion based on automatic differentiation and differentiable dynamic time warping}},
  author={Liu, Yingchang and Tang, Jie and Tang, Zhengwei and Sun, Chengyu},
  journal={Journal of Geophysics and Engineering},
  volume={20},
  number={3},
  pages={549--564},
  year={2023},
  publisher={Oxford University Press}
}

@inproceedings {TensorFlow,
author = {Mart{\'\i}n Abadi and Paul Barham and Jianmin Chen and Zhifeng Chen and Andy Davis and Jeffrey Dean and Matthieu Devin and Sanjay Ghemawat and Geoffrey Irving and Michael Isard and Manjunath Kudlur and Josh Levenberg and Rajat Monga and Sherry Moore and Derek G. Murray and Benoit Steiner and Paul Tucker and Vijay Vasudevan and Pete Warden and Martin Wicke and Yuan Yu and Xiaoqiang Zheng},
title = {{TensorFlow}: A System for {Large-Scale} Machine Learning},
booktitle = {12th USENIX Symposium on Operating Systems Design and Implementation (OSDI 16)},
year = {2016},
isbn = {978-1-931971-33-1},
address = {Savannah, GA},
pages = {265--283},
url = {https://www.usenix.org/conference/osdi16/technical-sessions/presentation/abadi},
publisher = {USENIX Association},
month = nov
}
